\documentclass[lettersize,journal]{IEEEtran}
\usepackage{amsmath}
\usepackage{multirow}
\usepackage{booktabs}
\usepackage{tabularx}
\usepackage{multicol}
\usepackage{bbding}
\usepackage{array}
\usepackage[table]{xcolor}
\usepackage{graphicx}
\usepackage{subfig}
\usepackage{soul}
\usepackage{wasysym}
\usepackage{lipsum}
\usepackage{array}
\usepackage{pifont}
\usepackage{tikz}
\usepackage{bigdelim}
\usepackage{algorithm,algpseudocode}
\usepackage{hyperref}
\usepackage{makecell}

\newcommand{\cmark}{\ding{51}}%
\newcommand{\xmark}{\ding{55}}%

\algtext*{EndWhile}
\algtext*{EndIf}
\algtext*{EndFor}
\algtext*{EndProcedure}

\newcommand{\ourtool}{\textit{Ambusher}}

\newcommand*\circled[1]{\tikz[baseline=(char.base)]{\node[shape=circle,draw,inner sep=0.3pt] (char) {#1};}}

\begin{document}

\title{\textsl{Ambusher}: Exploring the Security of Distributed SDN Controllers through Protocol State Fuzzing}

\author{Jinwoo Kim, Minjae Seo, Eduard Marin, Seungsoo Lee, Jaehyun Nam, and Seungwon Shin
\thanks{Jinwoo Kim is affiliated with the School of Software at Kwangwoon University, Seoul, South Korea (e-mail: jinwookim@kw.ac.kr).}%
\thanks{Minjae Seo is affiliated with ETRI, Daejeon, South Korea (e-mail: ms4060@etri.re.kr).}%
\thanks{Eduard Marin is affiliated with Telefonica Research, Barcelona, Spain (eduard.marinfabregas@telefonica.com).}%
\thanks{Seungsoo Lee is affiliated with the Department of Computer Science \& Engineering at Incheon National University, Incheon, South Korea (e-mail: seungsoo@inu.ac.kr).}%
\thanks{Jaehyun Nam is affiliated with the Department of Computer Engineering at Dankook University, Yongin, Gyeonggi-do, South Korea (e-mail: jaehyun.nam@dankook.ac.kr).}%
\thanks{Seungwon Shin is affiliated with the School of Electrical Engineering at KAIST, Daejeon, South Korea (e-mail: claude@kaist.ac.kr).}%
\thanks{\textit{This is the author’s accepted manuscript of the article published in IEEE Transactions on Information Forensics and Security}, Vol.~19, pp.~6264--6279, May~2024. DOI: \url{https://doi.org/10.1109/TIFS.2024.3402967}}
}


\maketitle

\begin{abstract}
Distributed SDN (Software-Defined Networking) controllers have rapidly become an integral element of Wide Area Networks (WAN), particularly within SD-WAN, providing scalability and fault-tolerance for expansive network infrastructures. However, the architecture of these controllers introduces new potential attack surfaces that have thus far received inadequate attention. In response to these concerns, we introduce \ourtool{}, a testing tool designed to discover vulnerabilities within protocols used in distributed SDN controllers. \ourtool{} achieves this by leveraging \emph{protocol state fuzzing}, which systematically finds attack scenarios based on an inferred state machine. Since learning states from a cluster is complicated, \ourtool{} proposes a novel methodology that extracts a single and relatively simple state machine, achieving efficient state-based fuzzing. Our evaluation of \ourtool{}, conducted on a real SD-WAN deployment spanning two campus networks and one enterprise network, illustrates its ability to uncover 6 potential vulnerabilities in the widely used distributed controller platform.
\end{abstract}

\section{Introduction}

\IEEEPARstart{O}{ver} the last few years, Software-Defined Networking (SDN) has gained significant attention from academia and industry, and it is currently being used in data center, telco, and enterprise environments~\cite{jain2013b4,hong2018b4,cord_web}. The SDN paradigm advocates for decoupling the network's intelligence (i.e., control plane) from the data forwarding functionality of network devices (i.e., data plane), placing the network's intelligence into a centralized SDN controller whose functionalities can be extended via SDN applications (i.e., application plane), and using a set of standard application programming interfaces (APIs)---commonly known as Northbound and Southbound interfaces---to facilitate communication between the planes. With this architecture, SDN provides considerable benefits to network operators including centralized network control and management as well as greater network programmability.

In the early days of SDN, it was thought that the SDN controller would be a single point of failure that could bring serious issues in terms of reliability, security, latency or scalability~\cite{yu2010scalable,shin2013attacking,skowyra2018effective,curtis2011devoflow}. However, besides the Northbound and Southbound interfaces, SDN architecture also contains \emph{East-West interfaces} to support communication between controller instances. Leveraging these interfaces, network operators often rely on a cluster of SDN controllers (i.e., one main controller and various replicas of it) to develop the control plane functionalities within the network. This way, if the main SDN controller fails or crashes, any of the replicas can immediately take over, leading to more robust and reliable networks. Likewise, to reduce latency and achieve better scalability, network operators typically divide large networks into various sub-networks, each managed by a different SDN controller. In such a case, the SDN controllers can even be located far away from each other, forming so-called Software-Defined Wide Area Networks (SD-WAN). Two prominent examples of SD-WANs are Google's B4~\cite{jain2013b4,hong2018b4} and Microsoft's SWAN~\cite{hong2013achieving}, which are used to interconnect their data centers distributed across the globe.


Despite its significant advantages, SDN brings new security challenges and attack vectors. Several researchers have demonstrated that adversaries can launch attacks against the application~\cite{xiao2020unexpected,ujcich2018cross,ujcich2020automated,xu2017attacking}, control~\cite{skowyra2018effective,hong2015poisoning,marin2019depth,shin2014rosemary,wang2015floodguard} and data planes~\cite{porras2015securing,cao2020match}. From the attacks proposed so far, those that target the controller are the most dangerous, given that the controller can be seen as the network’s brain. While there exists a wide range of attacks against SDN controllers~\cite{xu2017attacking,el2016sdnracer,cao2019crosspath,ujcich2017attain,skowyra2018effective}, the proposed attacks so far only consider networks with a \emph{single controller} and are launched via the Northbound or Southbound interfaces.

Unfortunately, no work has yet investigated the security of the protocols used in the East-West interfaces for a cluster of SDN controllers to communicate. As East-West protocols are used to perform critical functionalities within the controllers' cluster, such as choosing the leader SDN controller, selecting the controller that controls each networking device, or enforcing the network's policy, their security is crucial for the correct functioning of the network. If adversaries discover and exploit vulnerabilities in any East-West protocols, they could gain control of the controllers' cluster and execute attacks to disrupt the network, obtain sensitive network information, or poison the network's state. To make matters worse, the consequences of such attacks can be more significant than those executed against a single SDN controller. For example, in the SD-WAN case, adversaries located in one sub-network could perform remote attacks against other sub-networks far away from them~\cite{bustamante2021comparative,navarro2023software,seo2022heimdallr}.

As such, the security threats and risks in SDN networks with \emph{multiple controllers} remain unexplored to date. In this regard, thoroughly investigating security issues in East-West interfaces is a timely and challenging problem that can help network operators build a cluster resistant to security attacks. While many tools~\cite{canini2012nice,scott2014troubleshooting,mahajan2016jury,xu2017attacking,lee2017delta,jero2017beads,ujcich2017attain,dixit2018aim,lee2020audisdn,li2022spider,kim2023intender} have been proposed for detecting attacks within SDN environments, they are not suited for distributed SDN for various reasons. First, as modern distributed controllers involve diverse East-West protocols, existing tools require significant efforts to uncover vulnerabilities in such complex scenarios. Second, unlike the SDN Southbound interface (i.e., OpenFlow~\cite{openflow_web}), protocols used in East-West interfaces involve complicated states as many nodes within a cluster exchange multiple types of messages. However, existing tools cannot generate adequate test cases as they are not aware of such states of a cluster. While several tools~\cite{lee2017delta,lee2020audisdn,kim2023intender,ujcich2017attain} utilize a state machine for fuzzing, it is manually constructed, which is challenging to apply in distributed controllers.

Motivated by this problem, in this paper, we design and implement \emph{Ambusher}, which conducts \emph{protocol state fuzzing} for distributed SDN controllers. Protocol state fuzzing~\cite{de2015protocol,fiterau2020analysis} is a method that infers a target system's state machine and leverages it to identify unknown vulnerabilities or abnormal cases effectively. To apply this technique in a distributed controller environment with complex states, we propose a \emph{node-to-cluster} model that simplifies the learning process by treating a cluster as a single entity. Additionally, we introduce a fuzzing algorithm that utilizes the inferred state machine to generate acceptable message sequences and mutate them to uncover potential attack scenarios. We verified the feasibility of \ourtool{} by testing it on ONOS~\cite{berde2014onos}, a popular distributed SDN controller. Our evaluation, conducted on an SD-WAN testbed---which spans two campus networks and one enterprise network with an ONOS cluster---revealed 6 real vulnerabilities. We reported these disclosed vulnerabilities to the corresponding vendor and obtained CVEs (Common Vulnerability Exposures).

\noindent\textbf{Contributions.} Our contributions are summarized as follows:

\begin{itemize}
    \item We propose a learning methodology to extract a single and relatively simple protocol state machine from a distributed SDN cluster whose state machine is unknown.
    \item We design and implement \ourtool{} that conducts state-aware fuzzing by systematically producing message sequences from an inferred state machine to discover valid attacks.
    \item We evaluate \ourtool{} in an ONOS SDN cluster built upon an SD-WAN testbed and disclose 6 potential vulnerabilities.
    \item To the best of our knowledge, we are the first to investigate the security of the protocols being used in the East-West interfaces in distributed SDN controllers.
\end{itemize}

\begin{figure}[t]
    \centering
    \includegraphics[width=\linewidth]{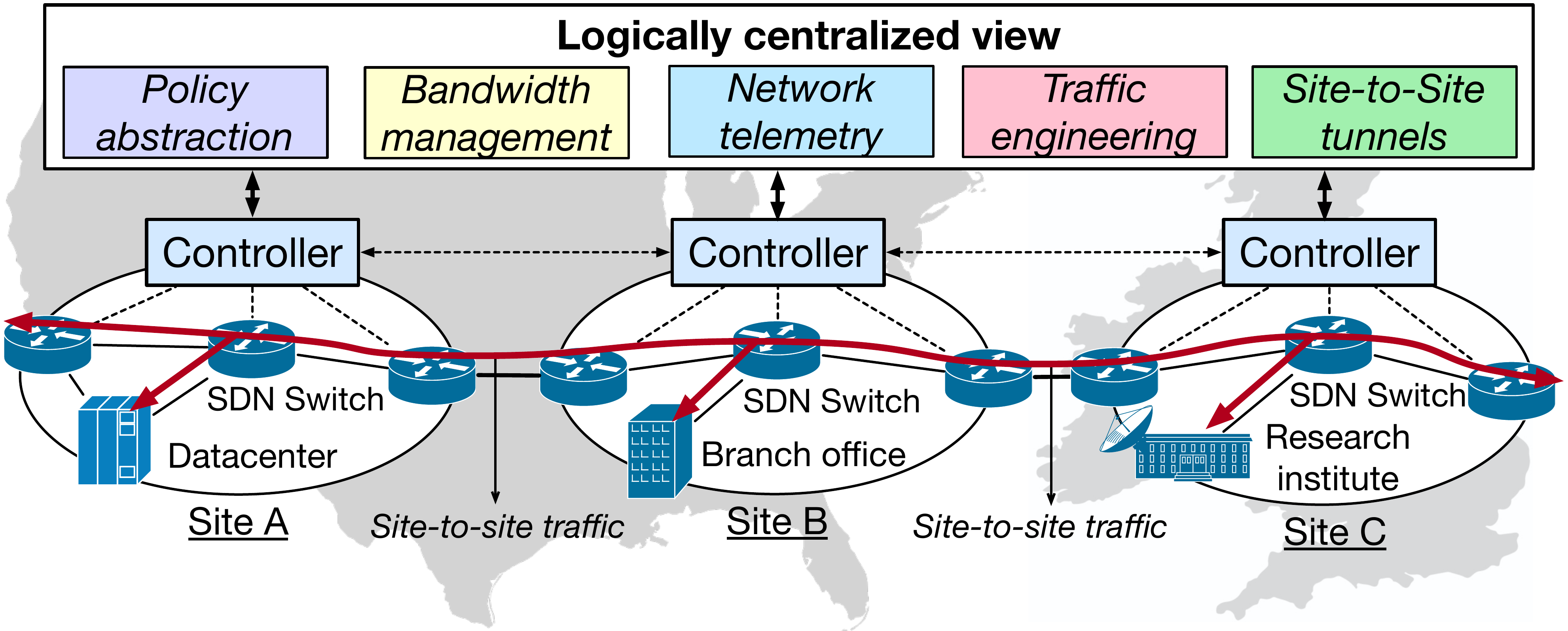}
    \caption{Distributed SDN controllers deployed in geographically different areas for building SD-WAN.}
    \label{fig:sdwan_example}
\end{figure}

\section{Background}
\label{sec:background}

In this section, we provide the necessary context to understand the architecture of an SDN cluster comprising multiple controllers.

\subsection{Distributed SDN Controllers}

Distributed SDN controllers were initially introduced within a single network to overcome limitations associated with a single controller, which was prone to a single point of failure and scalability challenges. The solution involves deploying multiple controller replicas, referred to as \emph{nodes}, which share states to form a \emph{cluster}. This cluster ensures the continuous operation of the control plane even in the event of a failure. Furthermore, distributing control-plane workloads across these nodes enhances overall performance and scalability. Well-established SDN controllers widely adopted in enterprise and telecommunication networks, such as ONOS~\cite{onos_web} and OpenDaylight~\cite{odl_web}, adhere to this distributed architecture.

Meanwhile, a cluster can be employed across multiple networks situated in geographically diverse areas, a concept known as Software-Defined Wide Area Networks (SD-WAN), widely embraced by major vendors such as Google and Microsoft~\cite{jain2013b4,hong2018b4,hong2013achieving}. Unlike the single network scenario, in SD-WAN, multiple nodes are strategically deployed to different physical locations. Each controller manages its network while synchronizing states to construct a logically centralized view. This configuration facilitates the seamless implementation of traffic engineering and optimization across physically distant areas, addressing a well-recognized challenge in WAN~\cite{yang2019software}. Fig.~\ref{fig:sdwan_example} provides an illustrative example of such a cluster deployment.

\subsection{Cluster Internal Architecture}
\label{subsec:protocol}

\begin{figure}[t]
    \centering
    \includegraphics[width=.9\linewidth]{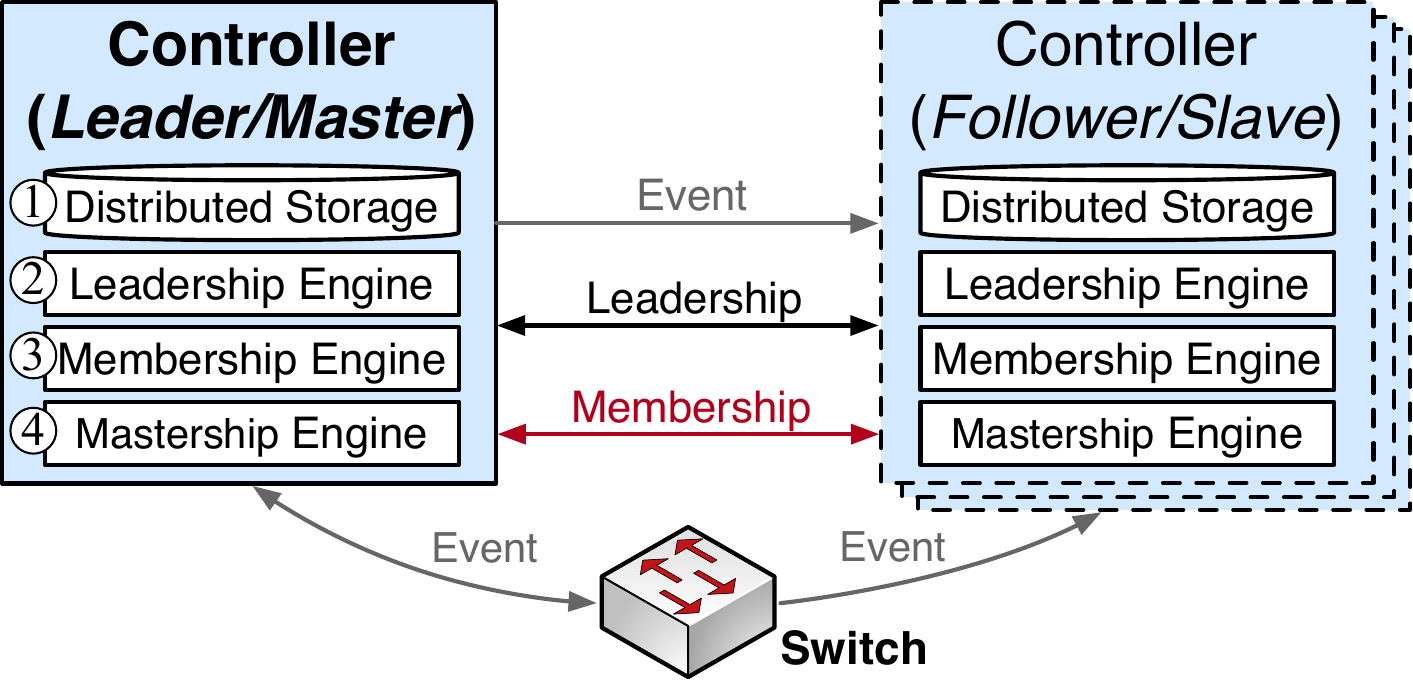}
    \caption{A general architecture of distributed SDN controllers.}
    \label{fig:cluster_architecture}
\end{figure}

Fig.~\ref{fig:cluster_architecture} illustrates a general architecture of an SDN cluster. It consists of the following four main components, namely: \circled{1}~\emph{distributed storage} for managing global network states, \circled{2}~\emph{leadership engine} for electing the cluster's leader, \circled{3}~\emph{membership engine} for periodically checking the aliveness of cluster nodes, and \circled{4}~\emph{mastership engine} for determining which controller manages the network devices (e.g., switches) within a given network segment. 
In what follows, we elaborate on how each of the previous building blocks within SDN clusters works.

\noindent\textbf{1) Distributed Storage.} One of the key goals of any SDN cluster is to maintain a consistent storage view among the cluster nodes~\cite{jain2013b4,berde2014onos,katta2015ravana}. For this, nodes must synchronize their storage with other nodes every time a new event (e.g., a change in the topology) occurs within their network segment. The way this is done is specified in consistency policies and depends on the type of event. For example, suppose control-plane events, such as those involving leadership/mastership within the SDN cluster, are detected. In that case, nodes must immediately notify the rest of the nodes, ensuring \emph{strong consistency}. On the contrary, other events, such as those triggered by network devices within a network segment, can be loosely synchronized, providing \emph{eventual consistency}.

\noindent\textbf{2) Leadership Engine.} In any SDN cluster, there is always a node that acts as a \emph{leader} for a given time duration, while the rest of the cluster nodes are known as \emph{followers}. During the leader election, all cluster nodes compete to be selected as the leader; as this procedure occurs periodically, the role of the leader can change over time. The leader node is in charge of replicating all network events to the follower nodes and keeping track of any changes produced in the storage of any of the followers. Instead, follower nodes only receive replicated states from the leader node, thus serving a backup role. Raft~\cite{ongaro2014search} is one of the most widely used algorithms for choosing a cluster leader. Its simplicity makes popular distributed controllers employ Raft dominantly (e.g., ONOS~\cite{berde2014onos}, OpenDaylight~\cite{medved2014opendaylight}).

\begin{figure}[t]
    \centering
    \includegraphics[width=\linewidth]{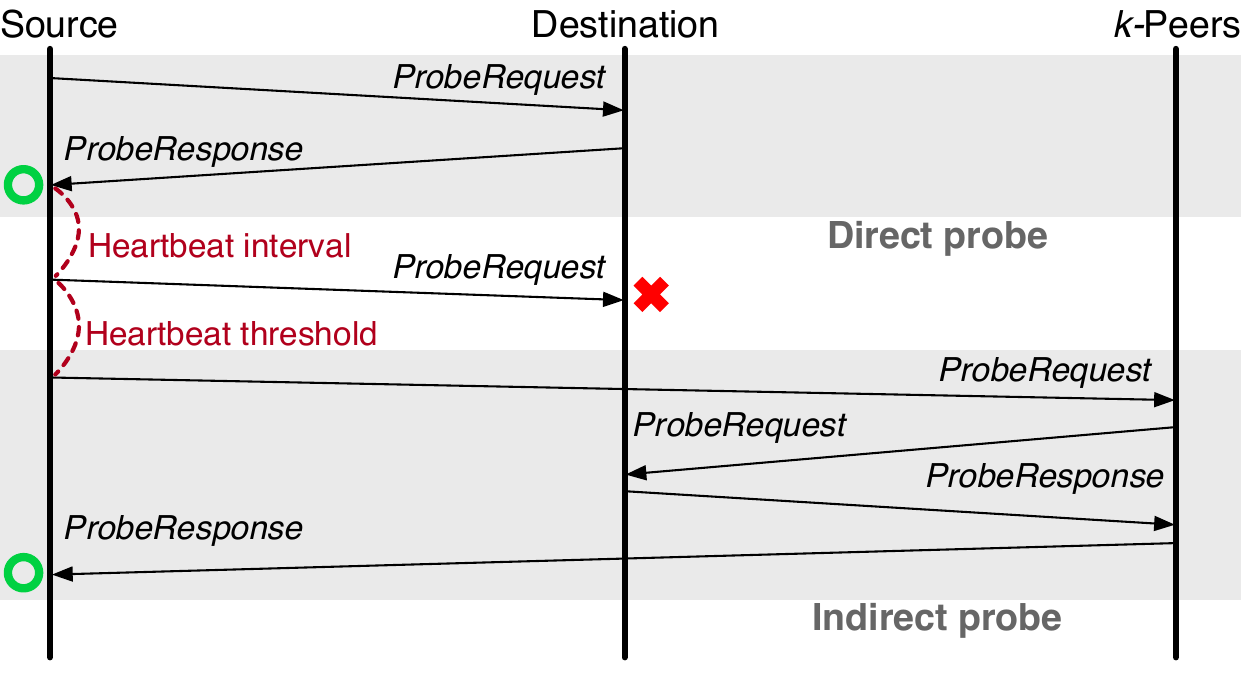}
    \caption{A sequence diagram of SWIM protocol~\cite{das2002swim}.}
    \label{fig:swim}
    \vspace{-0.1in}
\end{figure}

\noindent\textbf{3) Membership Engine.} Keeping nodes informed about the other nodes’ status is crucial to avoid Byzantine failures in distributed systems~\cite{lamport2001paxos}. To that end, nodes periodically check the aliveness of the other nodes by sending heartbeat messages to them. To carry out this task without introducing a significant overhead, one can employ an advanced probing solution like the one used by the SWIM or heartbeat protocols~\cite{onos_atomix_slide}. The primary objective driving these protocols is to enable each node to periodically select a subset of other nodes within the system, either randomly or based on a specified heuristic, for monitoring purposes. For example, in Fig.~\ref{fig:swim}, the source node first directly sends a \emph{ProbeRequest} message to the destination node. If the source node does not receive a \emph{ProbeResponse} within a certain heartbeat threshold, then the source node asks $k$ number of nodes to probe the destination node indirectly.

\noindent\textbf{4) Mastership Engine.} 
The mastership engine designates the node that controls network devices like switches. Through a controller-specific mastership election mechanism, each network device is assigned a unique \emph{master} node. This designated node holds write permissions, enabling it to modify the forwarding rules of the switch using a Southbound protocol like OpenFlow~\cite{openflow_web}. In contrast, the remaining nodes are confined to read-only access. Should a master node encounter a failure, the mastership is transferred to one of the \emph{slave} nodes through a re-election mechanism, dependent on the specific controller's protocol~\cite{odl_mastership,onos_mastership}.

\section{Problem Statement}

In this section, we first motivate the need to design a new testing framework for a cluster of SDN controllers. Next, we outline the main technical challenges and introduce our approach to identify potential vulnerabilities in the protocols being used in the East-West interfaces.

\subsection{Motivation} \label{sec:motivation}

So far, several testing tools have been proposed to systematically test SDN systems for attacks originating from the Northbound and Southbound interfaces. For example, Lee~\emph{et al.} introduced DELTA~\cite{lee2017delta}, a framework based on black-box fuzzing for automatically discovering vulnerabilities in the Northbound and Southbound interfaces (i.e., OpenFlow) of SDN controllers under different SDN deployments and threat models. DELTA randomizes sequences of APIs or OpenFlow packets before sending them to the controller and then analyzes the controller's response to each packet sequence. Following this work, Jero~\emph{et al.} presented BEADS~\cite{jero2017beads}, a similar testing framework for automatically uncovering vulnerabilities in SDN controllers triggered by malicious switches and hosts via the Southbound interface. Similarly to DELTA, BEADS utilizes blackbox fuzzing to identify vulnerabilities in SDN controllers; however, BEADS utilizes a more advanced fuzzer aware of protocol message formats and semantics, achieving higher test coverage. Another popular testing framework for SDN is the solution proposed by Ujcich \emph{et al.}~\cite{ujcich2017attain}. They proposed ATTAIN, a general attack injection framework that takes a testing specification (e.g., a set of attacks, the attacker capabilities, and network topology) as inputs from a network operator and reports how each of the defined attacks manifest in a given SDN controller.

Unfortunately, none of the existing tools is suitable for discovering potential weaknesses in the East-West protocols used by a cluster of SDN controllers to communicate with each other. This limitation is due to two main reasons: First, previous works did not put sufficient effort into retrieving the state machine of the analyzed protocol, a fundamental task in identifying potential attacks. Their methods to discover potential vulnerabilities are primarily based on a proxy that either modifies packets or randomizes a sequence of packets before sending them to the controller in an attempt to trigger corner cases, which can potentially lead to undesirable outcomes. However, as previous works did not consider the state the controller is in when receiving the packets, the identified attack test cases can contain many false positives and false negatives. Second and even more importantly, all previous tools considered only a single SDN controller and focused on one protocol only (e.g., OpenFlow). This limited scope renders existing tools unsuitable for discovering potential vulnerabilities in SDN clusters containing multiple controllers and protocols.

\subsection{Technical Challenges}

Learning a state machine is known as an efficient way to find vulnerabilities in protocol implementations~\cite{fiterau2020analysis,de2015protocol}. However, applying this method to an SDN cluster presents several challenges, as detailed below. Note that C1-C3 are associated with building a simple yet representative protocol state machine, while C4-C5 concern the identification of potential attacks or abnormal behaviors derived from the inferred state machine.

\noindent\textbf{C1) Need to infer the state machine of a cluster collectively.} SDN clusters typically involve multiple protocols to manage their internal components, which implies that any protocol can also affect the state transitions of the others. For example, whenever a \emph{ProbeResponse} is sent from a previously unknown node, the leader-election component leaves its current state and starts interacting with the new node. Hence, separately analyzing each protocol is insufficient to understand the security of the entire cluster.

\noindent\textbf{C2) Need to infer a simple state machine.} Each SDN controller within the cluster generates their own set of messages (with a broad range of message headers) to communicate with other nodes. In this context, learning protocol states without effectively pruning the negligible state will produce many states unnecessarily, commonly known as the state explosion. This approach is undesirable since having a complex protocol state machine would make a testing tool generate many test cases that require significant testing time for discovering attacks. Additionally, a complicated state machine prevents network operators from understanding cluster behavior.

\noindent\textbf{C3) Need to consider cluster synchronization.} An SDN cluster typically requires all controllers to be kept online because aliveness is crucial for many components, such as leader or mastership election. For example, in ONOS, a controller periodically sends keep-alive messages to other nodes. If a reply is not received for a certain threshold, that node is determined dead, reverting to an initial state. Thus, keeping the periodic interaction between a tester and target cluster is necessary. This approach ensures that the target cluster remains in an inferred state, facilitating the reproducibility of a discovered attack using the same message sequence.

\noindent\textbf{C4) Lack of an automated state fuzzing methodology.} Existing protocol state fuzzing methodologies~\cite{fiterau2020analysis,de2015protocol} rely extensively on manual analysis for vulnerability discovery.
Thus, network operators must examine the inferred state machine against a specification or ground-truth state machine to detect attacks or abnormal behavior. However, this manual process demands significant time commitment from operators and introduces challenges in efficiently reproducing attacks. Therefore, an automatic method is required to discover or generate attacks based on the inferred state machine of an SDN cluster.

\noindent\textbf{C5) Lack of ground truth for the state machine.} Distinguishing between legitimate and malicious behavior by inspecting the obtained protocol state machine is challenging because there is no ground truth on how the cluster’s protocol state machine should ideally be. If that existed, one could identify potential attacks by exploring how inconsistencies between the ideal and obtained protocol state machine could be exploited to carry out attacks.

\subsection{Our Approach}

To tackle C1 and C2, we adopt an abstraction approach by treating the entire set of nodes as a unified entity to construct a straightforward learning model. In particular, we treat a target cluster as a single node by focusing on the interaction with a leader node. Given the leader-centric event synchronization, this strategy leverages the observation that the leader node manages the majority of messages. Additionally, we streamline the input space of message headers by concentrating on a minimal set capable of triggering deterministic state transitions. To this end, we empirically select messages likely to induce state transitions by analyzing the controller source code. Consequently, network operators can derive a \emph{single and relatively simple protocol state machine} to facilitate reasoning about the security of the entire SDN cluster.

To tackle C3, we develop a \emph{state-aware learner} capable of identifying synchronization signals transmitted by a cluster. This learner responds with appropriate messages to maintain a valid state. It is important to note that this implementation necessitates the incorporation of protocol parsers specifically tailored to synchronization processes. To that end, we design a controller-specific proxy to generate synchronization messages needed for each state.

In response to C4, we formulate an algorithm for state machine fuzzing. This algorithm traverses an inferred state machine, extracting all conceivable message sequences originating from an initial state. Note that we define a \emph{message sequence} as a sequence of protocol messages that trigger state transitions of a cluster. By doing so, we automate the process of identifying and replicating attack scenarios, thereby minimizing operators' need for manual intervention.

In response to C5, we establish detection criteria to be employed when injecting diverse message sequences into the cluster. This approach involves continuous monitoring of alterations in both the controllers (e.g., detecting the deletion of applications) and the cluster (e.g., identifying changes in leadership or adding nodes). Furthermore, we gather information from various metrics, such as the CPU usage of each node, to unveil valuable insights for detecting attacks (e.g., elevated CPU usage in a node signaling a potential ongoing DoS attack).

\section{\ourtool{} Design}
In this section, we present system architecture of \ourtool{} with a workflow. We then discuss details of state learning and state fuzzing techniques for an SDN cluster.

\begin{figure}[t]
    \centering
    \includegraphics[width=.90\linewidth]{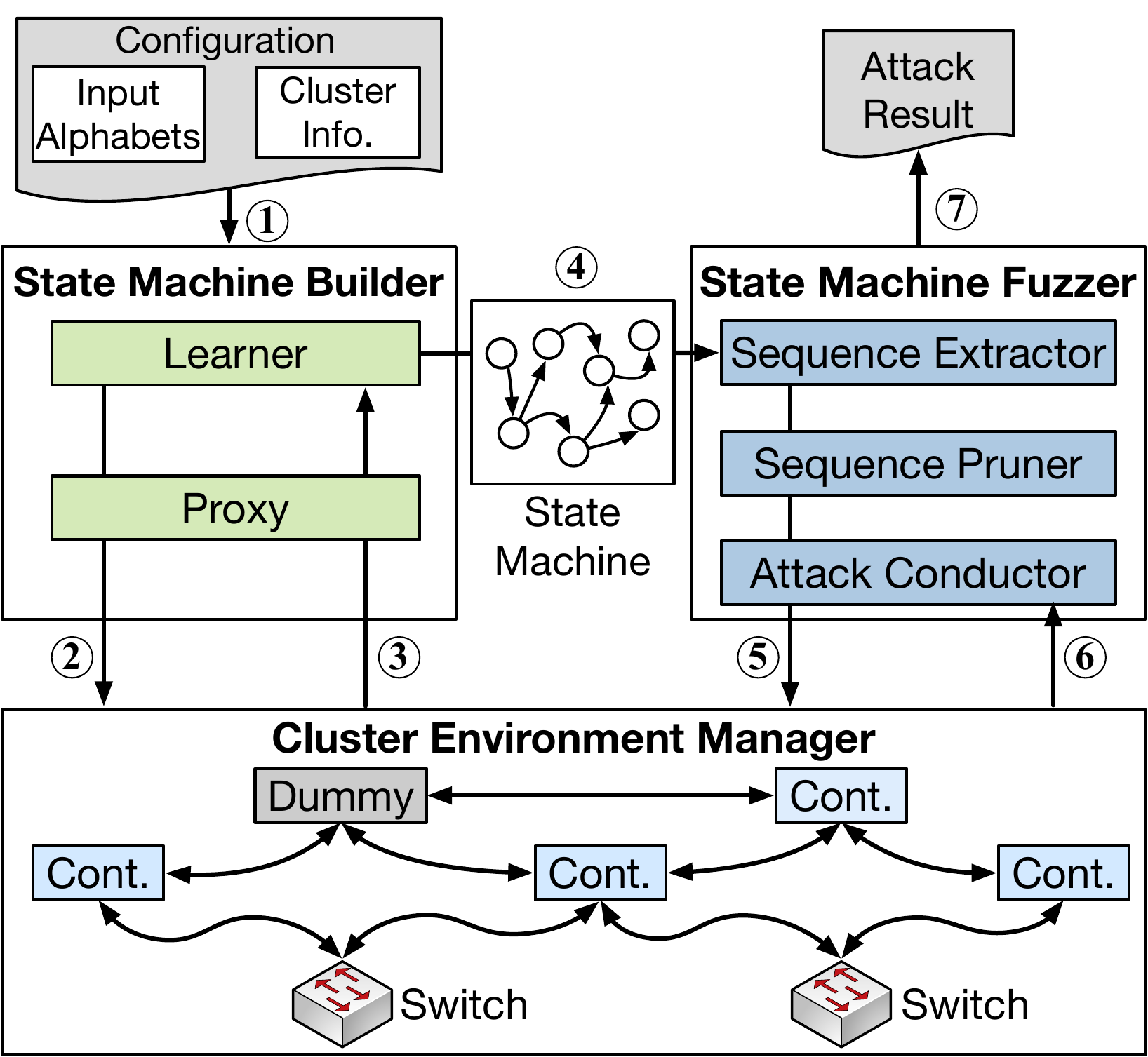}
    \caption{\ourtool{} system overview and workflow.}
    \label{fig:system_overview}
\end{figure}

\subsection{System Overview}

We aim to design \ourtool{} as a generic testing tool for distributed SDN controllers. To achieve this, a network operator can provide a testing configuration comprising input alphabets (i.e., abstract symbols of protocol messages) and cluster information (i.e., controller type, number of nodes, topology). At a high level, \ourtool{} consists of two phases: i) \emph{state machine learning} and ii) \emph{state machine fuzzing}. 
The former aims to learn a state machine from a specified cluster with predetermined input alphabets, while the latter seeks to identify attacks by mutating them. Note that \ourtool{} provides an initial set of input alphabets for popularly used East-West protocols, but they can be extended to incorporate other controllers (see \S\ref{sec:discussion}).

Fig.~\ref{fig:system_overview} illustrates the overall architecture of \ourtool{} and its workflow. It is composed of three main modules: (i) \emph{state machine builder}, (ii) \emph{state machine fuzzer}, and (iii) \emph{cluster environment manager}. In the state machine builder, the learner is responsible for inferring a state machine. \circled{1} For this purpose, it first takes a configuration from a network operator. 
\circled{2} The learner then generates queries used for learning states (\S\ref{sec:cluster_state}) and delivers them to the proxy, which in turn converts the queries into concrete protocol messages. These messages are then sent to the target cluster by a \emph{dummy node} created inside the cluster. \circled{3} When the proxy retrieves the response messages from the cluster, these messages are transferred to the learner oppositely to learn matched outputs. \circled{4} This loop continues as the learner discovers new states derived from new responses, halting only when no new states emerge, resulting in the creation of an inferred state machine.
\circled{5} The sequence extractor in the state machine fuzzer explores the state machine and chooses a message sequence that the cluster environment can accept. After the sequence pruner removes unnecessary inputs that do not affect state transitions, the attack conductor uses it as a seed. \circled{6} It iteratively randomizes the message sequence (\S\ref{sec:state_fuzzing}), executes it, and retrieves outputs from cluster logs. \circled{7} Finally, the state machine fuzzer yields the attack result, which is subsequently analyzed with several criteria for finding attacks 
(\S\ref{subsec:criteria}).

\begin{figure}[t]
    \centering
    \includegraphics[width=\linewidth]{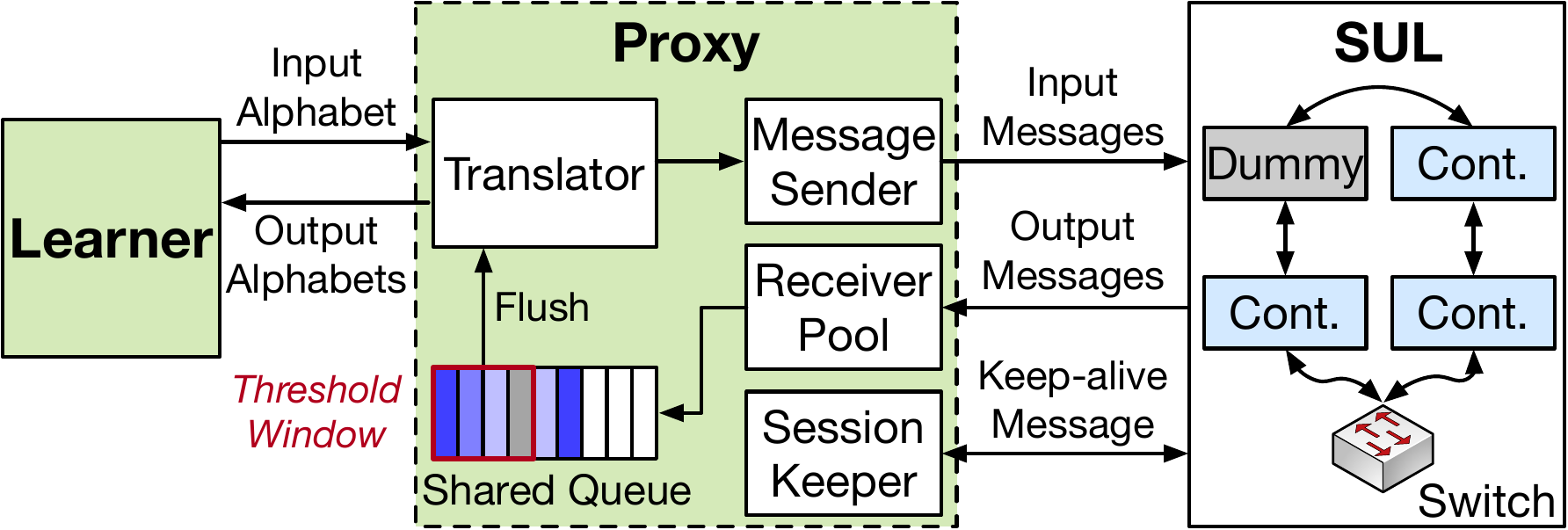}
    \caption{The workflow of the learning procedure in the state machine builder.}
    \label{fig:learning_setup}
\end{figure}

\subsection{Learning State Machine}
\label{sec:cluster_state}

Here, we elaborate on the details of the learning technique to infer the internal states of an SDN cluster. It is important to note that our goal is a more challenging task than the previous studies aimed at learning states or transitions of well-known protocols (e.g., TLS/DTLS ~\cite{de2015protocol,fiterau2020analysis}) because SDN East-West protocols do not have standard specification and its implementation is dependent to controller vendors. What is worse, protocol dependencies can be dramatically complicated and increased depending on the cluster size (e.g., number of nodes, configuration settings).

\noindent\textbf{Automata Learning.} To address this, we use \emph{automata learning}, which is a framework for systematically inferring a finite state machine (FSM) of a target system~\cite{steffen2011introduction}. It enables us to learn a simple and abstract FSM by interacting with the target system. Among many FSMs, the Mealy machine has been primarily used for protocol state fuzzing since it is well suited to understand protocol behavior due to its deterministic property---the state transition is determined by a unique input and state~\cite{de2015protocol,fiterau2020analysis}. Here, we call a Mealy machine if a state machine is an FSM whose outputs are determined by current states and inputs. In order to form a series of learning procedures, the framework is composed of two main concepts: (i) \emph{learner} and (ii) \emph{system under learning (SUL)}. The learner is responsible for inferring the Mealy machine of the given SUL, which is the target cluster environment in our case. During a learning process, the learner iterates the exploration and testing phases. In the exploration phase, predetermined symbols (i.e., input alphabets) are sent to the SUL to observe its responses (i.e., output alphabets). 
Once suitable responses are observed, the learner constructs a hypothesis model, a minimal Mealy machine whose states conform to the observation. In the testing phase, the hypothesis model is verified by finding whether or not there is a counterexample that violates it. If no one is found, the hypothesis model is accepted; otherwise, the model is refined. Those tasks are repeated until no counterexample is found from the model.

\noindent\textbf{Learning Model Design.} The alphabets are abstract symbols that SUL does not understand since they are not protocol messages. For this, it is necessary to have an intermediate \emph{proxy} that interprets the input alphabets into concrete messages. On the other hand, protocols in an SDN cluster typically have \emph{keep-alive} messages for aliveness checking---when a new node joins a cluster, existing members send those keep-alive messages to the node periodically. Whereas responding to the messages is inevitable to maintain a valid East-West session with the SUL, they are merely required to learn states due to uniformity. Furthermore, protocols use individual sessions to transmit/receive messages, and the communications run parallel among nodes. Given this concurrency, determining which output is derived from which input is challenging.

\begin{figure}[t]
    \centering
    \includegraphics[width=.70\linewidth]{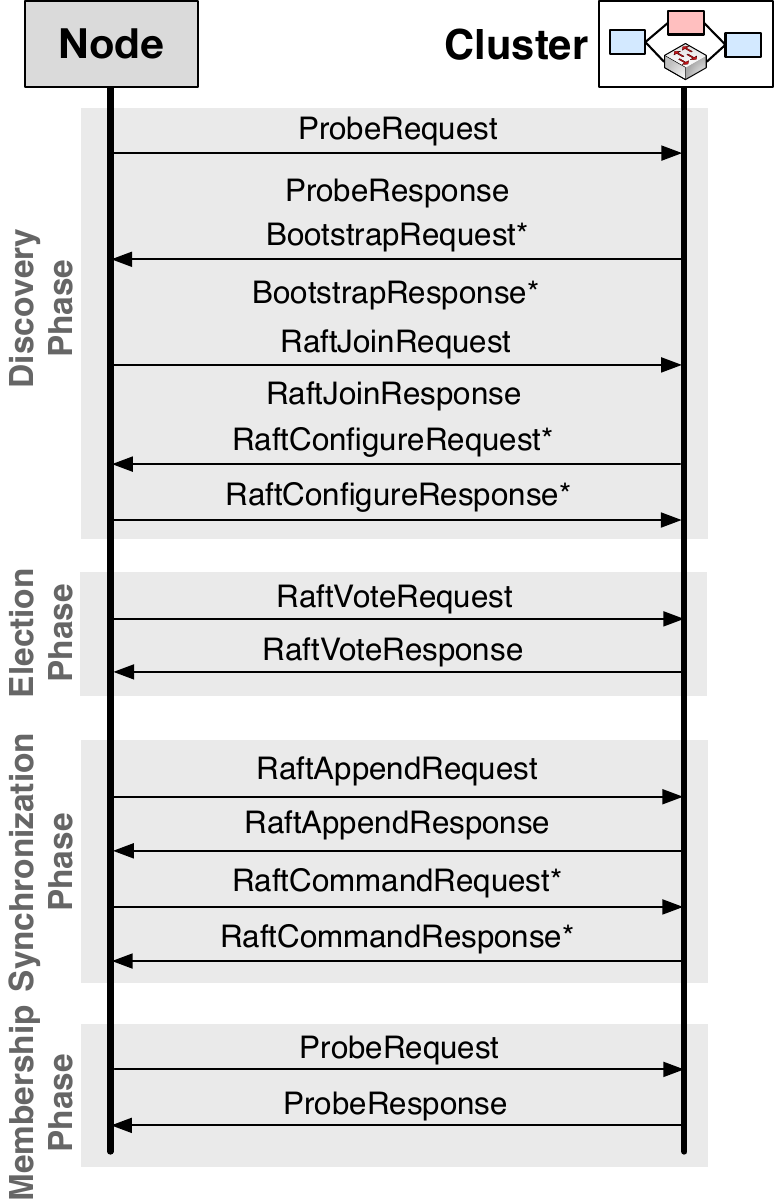}
    
    \footnotesize{$\ast$ ONOS/Atomix implementation-specific messages.}
    \caption{The abstract model for interactions between a dummy node and ONOS cluster.}
    \label{fig:protocol_model}
    \vspace{-0.1in}
\end{figure}

\begin{table}[t]
    \small
    \centering
    \caption{A list of ONOS cluster messages that are used for input/output alphabets for automata learning.}
    \begin{tabular}{l l l l}
        \toprule
        & \textbf{Alphabet} & \textbf{Shorthand} \\
        \midrule
        
        \ldelim\{{2}{0.1cm}[\parbox{0.2cm}{\rotatebox[origin=c]{90}{Membership}}] \ldelim\{{13}{0.1cm}[\parbox{0.2cm}{\rotatebox[origin=c]{90}{Discovery}}]
        
        & ProbeRequest($n$) & PReq($n$) \\
        & \multicolumn{2}{c}{$n\in\{n', n''\}$, $n' \in N$, $n'' \notin N$} \\
        & ProbeResponse($n$, $s$) & PRes($n$, $s$) \\
        & \multicolumn{2}{c}{$n \in N, s \in \{alive, dead\}$} \\
        
        & BootstrapRequest($N'$) & BReq($N'$) \\
        & \multicolumn{2}{c}{$N' \subseteq N$} \\
        & BootstrapResponse($N'$) & BRes($N'$) \\
        & \multicolumn{2}{c}{$N' \subseteq N$} \\
        & RaftJoinRequest($n$) & RJReq($n$) \\
        & \multicolumn{2}{c}{$n\in\{n', n''\}$, $n' \in N$, $n'' \notin N$} \\
        
        & RaftJoinResponse & RJRes \\
        
        & RaftConfigureRequest & RConReq \\
        
        & RaftConfigureResponse & RConRes \\
        
        \ldelim\{{4}{0.1cm}[\parbox{0.2cm}{\rotatebox[origin=c]{90}{Election}}]
        & RaftVoteRequest($n, t$) & RVReq($n, t$) \\
        & \multicolumn{2}{c}{$n \in N, t\in\{t_h, t_c\}, t_h > t_c$}  \\
        & RaftVoteResponse($v$) & RVRes($v$) \\
        & \multicolumn{2}{c}{$v\in\{approved, rejected\}$} \\
        
        \ldelim\{{4}{0.1cm}[\parbox{0.2cm}{\rotatebox[origin=c]{90}{Synchronization}}]
        & RaftCommandRequest($d$, $o$) & RComReq($d$, $o$) \\
        & \multicolumn{2}{c}{$d\in\{app,topo\}$, $o\in\{add,modify,remove\}$} \\ 
        & RaftCommandResponse & RComRes \\
        & RaftAppendRequest & RAReq \\
        & RaftAppendResponse & RARes \\
        
        & NoResponse & - \\
        \bottomrule
    \end{tabular}
    \label{tab:cluster_messages}
    \vspace{-0.1in}
\end{table}

We design the proxy to incorporate the considerations above, as shown in Fig.~\ref{fig:learning_setup}. When the \emph{translator} receives an input alphabet, it converts the symbol into a protocol message. Since there is currently no standard for East-West protocols, most distributed controllers use their custom protocol implementations. Hence, this conversion requires us to analyze the controller's source code to create concrete messages.
The concrete message is sent by the \emph{message sender} connected to the dummy node in SUL. The \emph{receiver pool} dynamically invokes creating a new thread in case a new channel is established with SUL. To process the messages that are sent concurrently, we use the \emph{shared queue}. If the sender forwards messages to SUL, threads in the receiver pool enqueue the received messages into the shared queue. Then the translator dequeues the messages within a threshold window (We use the same threshold with the heartbeat threshold of a cluster configuration.). The message sender maintains a logical clock and produces a timestamp when sending an input message. By checking this clock, the translator guarantees a correct message order. The \emph{session keeper} is used for answering the keep-alive messages to maintain East-West channels, but the received messages are not used for learning. As such, utilizing the proxy plays an essential role in bridging the gap between the learner and SUL by converting the symbol into a protocol message.

\noindent\textbf{Modeling Node-to-Cluster Interactions.} Understanding cluster behavior holistically should be preceded to avoid complex state learning in distributed environments. In general, the leader manages the cluster; thus, most messages are answered by the leader node. Given this fact, we devise an abstract model that illustrates unified interactions from the aspect of the \emph{node-to-cluster} relation, not the protocol-wise communications. Fig.~\ref{fig:protocol_model} depicts the message interactions that occur when a dummy node starts communicating with a target cluster from scratch. We analyze ONOS/Atomix~\cite{onos_web,atomix_web} as a representative SDN cluster that realizes the aforementioned distributed architecture/protocols. Below, we introduce four interaction phases with brief descriptions of protocol messages if not given in \S\ref{sec:background} (e.g., implementation-specific messages).

\textit{1) Discovery phase.} This phase aims to discover and join a target cluster as a legitimate member. Initially, the node sends \emph{ProbeRequest} that subsequently triggers the cluster to reply with \emph{ProbeResponse} and \emph{BootstrapRequest}. The latter contains configuration information, such as cluster members/protocols. The node sends with \emph{BootstrapResponse} that specifies its architectural information (e.g., protocols, nodes), and it also sends \emph{RaftJoinRequest} for joining the cluster as a Raft member. Subsequently, the cluster responds with \emph{RaftJoinResponse} and \emph{RaftConfigureRequest} that carries current Raft protocol information (e.g., term, leader).

\textit{2) Election phase.} Messages in this phase are mostly related to the leader election process in the Raft protocol. When joining a cluster, the node's default role is assigned as a follower. The node has its election timer, and when the timer expires, it attempts to promote to a leader by sending \emph{RaftVoteRequest} messages to the cluster. The node will become a leader if it receives \emph{RaftVoteResponse} messages that most nodes agree.

\textit{3) Replication phase.} The node joined in the cluster starts to synchronize events. When receiving \emph{RaftAppendRequest} that includes a commit message pushed by a leader, the node will send a \emph{RaftAppendResponse}, noticing it has received the request and executed the commit. The node can read and modify shared views (e.g., application, topology) in distributed storage using \emph{RaftCommandRequest}, and the cluster notifies the result with \emph{RaftCommandResponse}.

\textit{4) Membership phase.} A node periodically checks the aliveness of peer nodes based on membership protocols. When a SWIM protocol is used, a node randomly chooses a target node and sends \emph{ProbeRequest} that specifies its identifier to the cluster and receives a \emph{ProbeResponse} message.

\begin{figure*}[t]
    \centering
    \includegraphics[width=\textwidth]{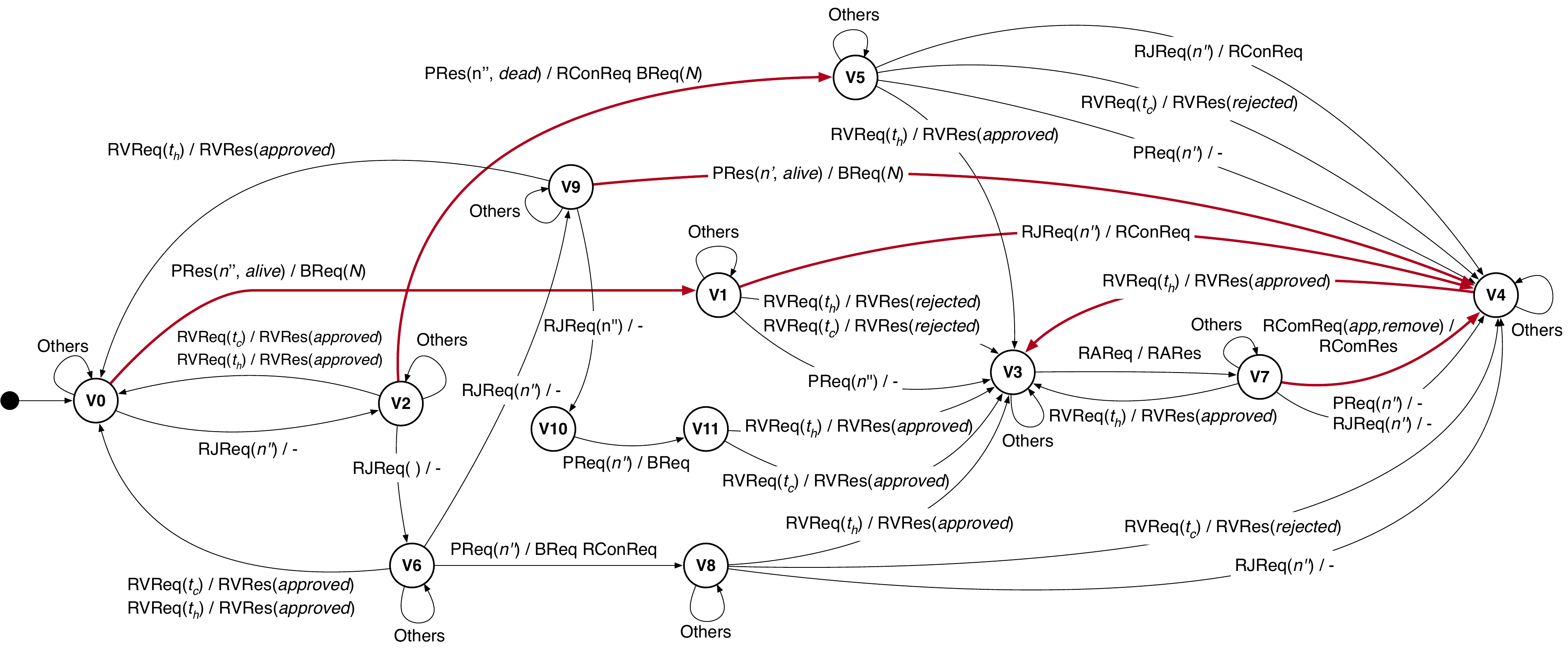}
    \caption{The automatically constructed Mealy machine of an ONOS cluster. The red arrows denote the state transitions that can be abused for attacking a cluster. Please refer to Table~\ref{tab:cluster_messages} for abbreviations.}
    \label{fig:state_machine}
    \vspace{-0.1in}
\end{figure*}

\noindent\textbf{Defining Alphabets.} 
Based on the message exchange model, we define alphabets used for learning a state machine as summarized in Table~\ref{tab:cluster_messages}. To make a state machine that incorporates diverse cases, we pick and choose the message \emph{parameters} that can affect the internal states of a cluster based on the manual analysis of source code (We discuss how this can be achieved in \S\ref{sec:discussion}.). 
For example, \emph{ProbeResponse($n,s$)} implies that the node $n$ is in the membership status $s$, which can be either $alive$ or $dead$ for answering \emph{ProbeRequest($n$)}. The \emph{BootstrapRequest($N'$)} and \emph{BootstrapResponse($N'$)} indicate that there are a set of nodes $N'$ currently configured in the cluster. The variable $N'$ can be a subset of the entire node set $N$. The \emph{RaftJoinRequest($n$)} denotes that a node $n$ joins a Raft protocol interaction, subsequently replied by \emph{RaftJoinResponse}. The variable $n$ can be a current member node $n'$ or a new node $n''$. The \emph{RaftConfigureRequest} indicates a configuration request for a Raft cluster, which is responded to by \emph{RaftConfigureResponse}. The \emph{RaftVoteRequest($n, t$)} denotes that a node $n$ attempts to be promoted to a leader with a term $t$ that can be a greater term $t_h$ or current term $t_c$. The \emph{RaftVoteResponse($v$)} is used for notifying a voting result that can be either $approved$ or $rejected$. The \emph{RaftCommandRequest($d,o$)} instructs a cluster to execute an operation $o$ that can be either $add$, $modify$, or $remove$ for a shared data $d$, and it can be important information, such as application, topology, or mastership information (denoted by $app$, $topo$, respectively). Fig.~\ref{fig:state_machine} illustrates an example of a state machine learned from an ONOS/Atomix cluster with those alphabets.

\begin{algorithm}[t]
\footnotesize
\caption{SDFS for Message Sequence Extraction}
\label{alg:dfs}
\begin{algorithmic}[1]
\Require 
\Statex A Mealy machine graph $\mathcal{G}=(\mathcal{V},\mathcal{E}, \mathcal{I}, \mathcal{O})$,
\Statex An initial state $v_0$
\Ensure 
\Statex A set of message sequences $\mathcal{M}$

\Procedure{Init}{$\mathcal{G}, v_0$}
        \State $\mathcal{S} \gets []$ \Comment Empty list
        \State $\mathcal{M} \gets \{\}$  \Comment Empty set
        \State Set all states in $\mathcal{G}$ as not visited
    	\State $\mathcal{M} \gets$ \Call{SDFS}{$\mathcal{G}, v_0, \mathcal{M}, \mathcal{S}$}
    	\State \Return $\mathcal{M}$
\EndProcedure
\Require 
\Statex A currently visited state $v$,
\Statex A subsequence that consists of states visited previously $\mathcal{S'}$

\Procedure{SDFS}{$\mathcal{G}, v, \mathcal{M}, \mathcal{S}_{pre}$}
        \For{$e \in \mathcal{G}.outgoingEdges(v)$, where $e = (v, w)$}
            \If{$w$ is not visited}
                \State Mark $w$ as visited
                \State $m \gets \mathcal{I}(e)$ \Comment{Get a message from a transition}
                \State $\mathcal{S}_{pre}.append(m)$ \Comment{Add the message to the sequence}
                \State $\mathcal{M} \gets \mathcal{M} \cup \mathcal{S}_{pre}$ \Comment{Add the subsequence to the set} 
                \State $\mathcal{S}_{post} \gets$ \Call{SDFS}{$\mathcal{G}, w, \mathcal{M}, \mathcal{S}_{pre}$} \Comment{Call SDFS recursively}
                \State $\mathcal{M} \gets \mathcal{M} \cup \mathcal{S}_{post}$ \Comment{Add the subsequence to the set}
            \EndIf
        \EndFor
        \State \Return $\mathcal{M}$
    	
\EndProcedure

\end{algorithmic}
\end{algorithm}


\subsection{State Machine Fuzzing}
\label{sec:state_fuzzing}

We now present a fuzzing technique that uses a state machine to produce test cases systematically. This stage aims to generate a set of message sequences that allow us to explore as many states as possible.

\noindent\textbf{State Machine Formalization.} To utilize the constructed Mealy machine for fuzzing, we first should formalize it with a suitable structure. A Mealy machine can be represented as a directed, multi-edged graph $\mathcal{G}=(\mathcal{V},\mathcal{E})$, where $\mathcal{V}$ denotes the states and $\mathcal{E}$ denotes the transitions labeled by corresponding input alphabets $\mathcal{I}$ and output alphabets $\mathcal{O}$. The $\mathcal{I}$ and $\mathcal{O}$ are functions that map a transition $e \in \mathcal{E}$ to message $m$.

\noindent\textbf{Pruning Transitions.} To reduce efforts for exploring states, we prune alphabets that do not affect transitions from the state machine. For example, a state machine can have a \emph{loop} that a state is connected with itself. Besides, keep-alive messages merely activate meaningful transitions such as \emph{ProbeRequest} and \emph{RaftAppendRequest}. We also exclude those transitions from the set of candidate message sequences, and they are denoted by \texttt{Others} in the state diagram.

\noindent\textbf{Message Sequence Extraction.} We want to explore all reachable states and generate possible message sequences that can be made from a state machine. For this, we propose an algorithm that extracts message sequences using graph depth-first search (DFS) as shown in Algorithm~\ref{alg:dfs}. The SDFS (State DFS) algorithm takes a Mealy machine graph $\mathcal{G}$ and initial state $v_0$ as inputs and yields a set of message sequences $\mathcal{M}$ as an output. 
SDFS initializes two variables $\mathcal{S}$ and $\mathcal{M}$ (lines 2 to 3). The former is used for storing messages extracted on visited states so far, and the latter is the algorithm's output, which will have the final set of message sequences in the end. At first, the algorithm marks all states as not visited and starts a traversal from an initial state $v_0$ (lines 4 to 5). When invoked, SDFS finds all outgoing edges (i.e., transitions) $e$ from the current state $v$, and checks whether or not the next state $w$ is visited (lines 8 to 9). If not, the algorithm marks the next state $w$ as visited and gets a message $m$ from a transition $e$ (lines 10 to 11). The message is added to the sequence $\mathcal{S}_{pre}$, and it is also added to the set $\mathcal{M}$ (lines 12 to 13). The SDFS is recursively invoked by having the next state $w$ and the pre-sequence $\mathcal{S}_{pre}$, and subsequently produces post-sequence $\mathcal{S}_{post}$ (lines 14). Finally, the message set $\mathcal{M}$ that includes pre- and post-sequences is generated (line 16). Note that the worst-case time complexity is $O(|\mathcal{V}|+|\mathcal{E}|)$ considering that the algorithm visits all nodes $\mathcal{V}$ for initialization (line 4) and iterates all transitions $\mathcal{E}$ (line 8) for traversing all adjacent states. Note that other operations take constant time like adding elements to a set or list (i.e., $O(1)$).

\noindent\textbf{Sequence/Message Randomization.} 
Injecting the same message sequence that always follows the state machine will not lead to attacks. To address this, \ourtool{} randomizes an extracted message sequence to make a target cluster in abnormal status. The primary fuzzing strategy of \ourtool{} is frequently randomizing message orders while minimally modifying message parameters. This strategy stems from the fact that randomizing the entire message parameters generates many test cases, often called a space explosion. Suppose that we want to randomize a message sequence $\mathcal{S}=(m_1, m_2, ... , m_n)$---extracted from Algorithm~\ref{alg:dfs}---that consists of $n$ messages. At each fuzzing stage, \ourtool{} chooses a message $m_i$, where $i$ is randomly chosen from the range $1 \leq i \leq n$. For this, \ourtool{} takes one of the following actions:
\begin{itemize}
    \item Duplicate the message $m_i$ several times.
    \item Remove the message $m_i$ from the sequence $\mathcal{S}$.
    \item Replace it with another message randomly picked from the sequence $\mathcal{S}$.
    \item If the message $m_i$ has an argument, change the argument to other valid ones (as defined in Table~\ref{tab:cluster_messages}).
\end{itemize}
This way, \ourtool{} can increase the probability of occurring abnormal events from a target cluster. 

\subsection{Detection Criteria}
\label{subsec:criteria}

After injecting the mutated message, assessing its potential to launch a successful attack on the target cluster is crucial. To accomplish this goal, we define \emph{detection criteria (DC)} aimed at identifying abnormal situations that pose risks to a cluster's confidentiality, integrity, and availability. These criteria are intricately crafted to detect and signal any deviations that may endanger the confidentiality of sensitive data, compromise the integrity of stored information, and disrupt the overall availability of controller nodes.

\noindent\textbf{DC1) Leaking Cluster Information.} Receiving a message that leaks information about a cluster indicates a potential threat. For example, \emph{BootstrapRequest} contains configuration information such as cluster members and protocols. Suppose that a malicious node can receive this message by injecting a certain message sequence (discovered with \ourtool{}) that allows a node to join a cluster. Then, it can learn this information and hence compromise the confidentiality of the cluster.

\noindent\textbf{DC2) Cluster Configuration Changes.} \ourtool{} periodically updates the global topology view to check node and link connection status. At the same time, it identifies the current leader in the cluster by parsing the updating protocol messages. Accordingly, it can compare the modified information with the original one. This criterion enables us to define the attacks aiming to compromise the cluster configuration integrity (i.e., topology, leadership).

\noindent\textbf{DC3) Cluster Status Changes.} The status of a cluster, including its applications, holds significant relevance in cluster management. In addressing this, \ourtool{} employs a linear search to update the app-list from the internal storage of each node within the cluster (e.g., persistent storage or in-memory storage). Subsequently, it scrutinizes the app-list information to detect any manipulations, allowing us to define attacks targeting the compromise of cluster status changes that violate integrity.

\noindent\textbf{DC4) Network Reachability Changes.} Another crucial indicator is end-to-end reachability between nodes. Specifically, \ourtool{} performs a pair-wise ICMP ping test to identify that all nodes in the cluster are verified. Therefore, it enables us to define attacks aiming to compromise the availability of network connections.

\noindent\textbf{DC5) Excessive Resource Usage.} The final indicator to warn of vulnerability is monitoring the RAM and CPU usage periodically used by the nodes in the cluster. \ourtool{} reports benign usage history in the storage to define any denial of service trial if excessive usage is detected from the monitoring. This metric enables us to identify attacks relating to network service availability.

\section{Experimental Environment}
\label{sec:exp}
\begin{figure}[t]
    \centering
    \includegraphics[width=.9\linewidth]{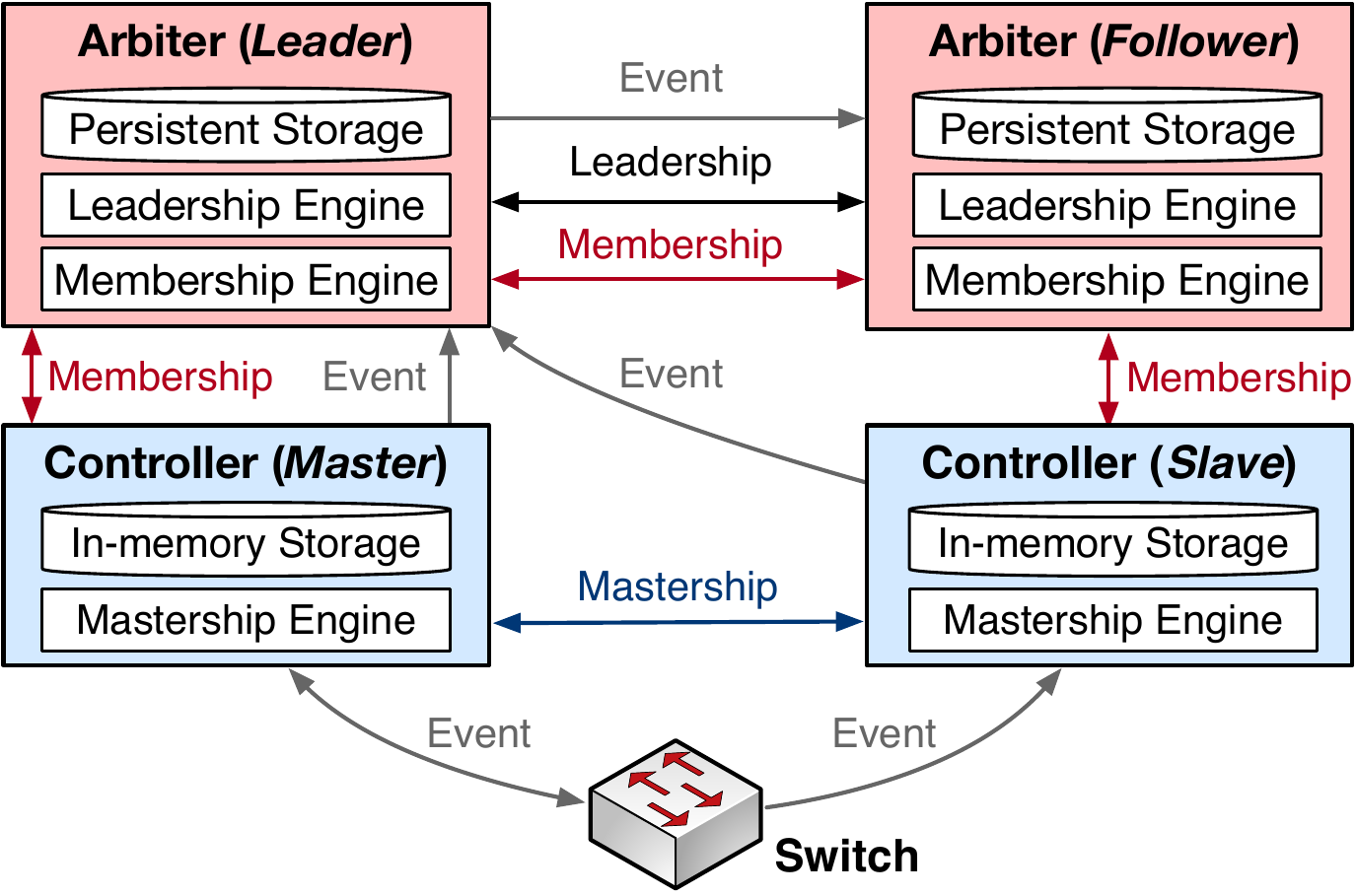}
    \caption{The architecture of an ONOS/Atomix cluster.}
    \label{fig:onos_architecture}
    \vspace{-0.10in}
\end{figure}

\noindent\textbf{Implementation.} We implemented a prototype of \ourtool{} based on Learnlib v0.14~\cite{learnlib_web}, which is an automata learning framework used for inferring states of a target system. To learn state machines of distributed controllers, we used the L$\ast$ algorithm~\cite{angluin1987learning} for the exploration phase and the W-method~\cite{chow1978testing} for the testing phase (\S\ref{sec:cluster_state}). For verifying the feasibility of \ourtool{}, we tested ONOS v2.4.0~\cite{onos_web}, one of the popular open-source SDN controllers widely used in practice~\cite{onos_att_web,riverbed_sdwan_web,trellis_web,verizon_web}. Note that in the recent ONOS architecture, some of the controller functionalities are realized in a separate node called \emph{Arbiter} to reduce the load of the cluster nodes (see Fig.~\ref{fig:onos_architecture}). Thus, a set of arbiter and controller nodes together forms a cluster. For this, we also used Atomix v3.1.5~\cite{atomix_web} for running arbiters and providing APIs for underlying cluster engines of ONOS controllers. To implement the \emph{proxy} in the state machine builder, we leveraged APIs provided by Atomix to generate cluster messages used in ONOS.

\begin{figure}[t]
    \centering
    \includegraphics[width=\linewidth]{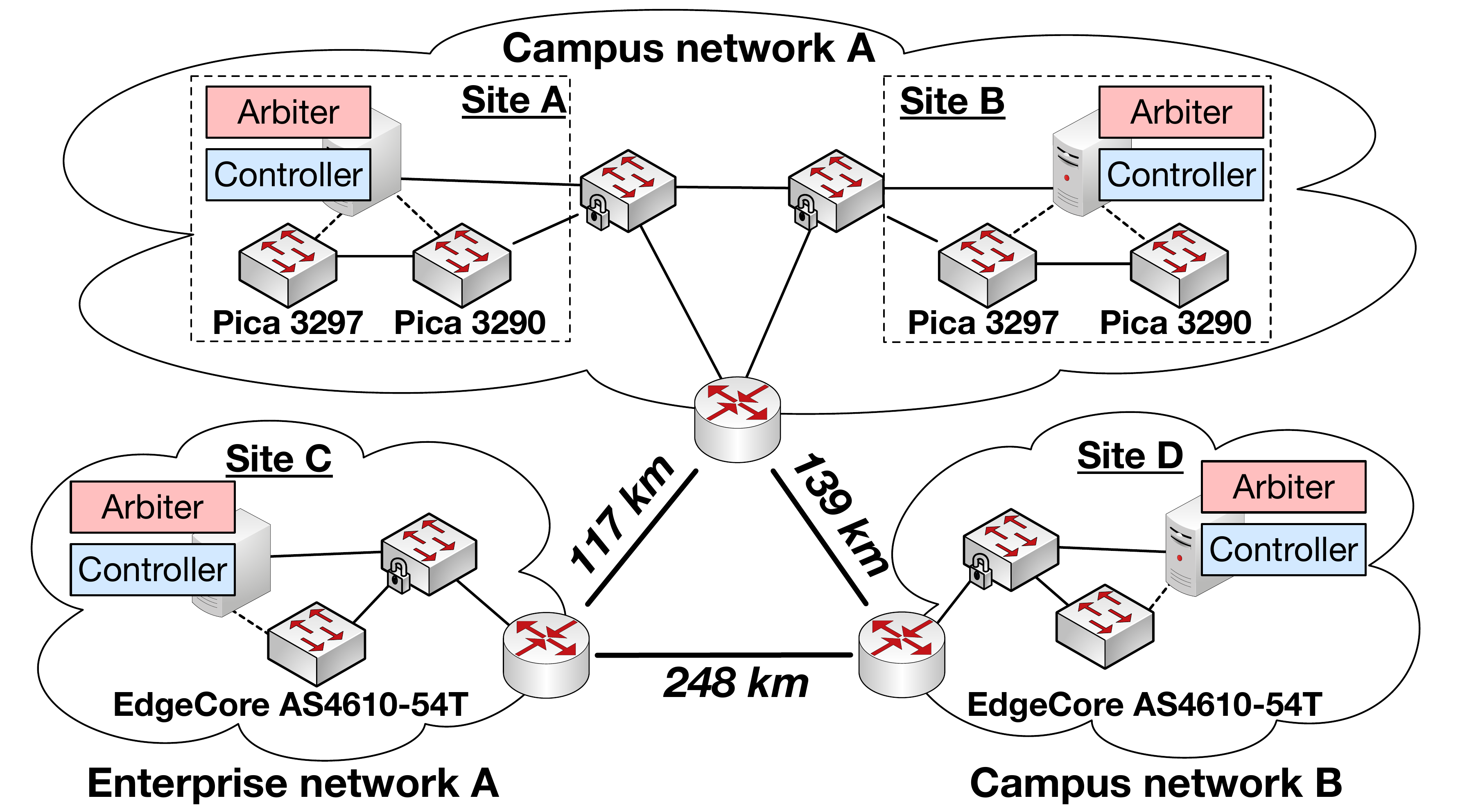}
    \caption{Private SD-WAN testbed architecture overview.}
    \label{fig:environment}
    \vspace{-0.10in}
\end{figure}

\noindent\textbf{SD-WAN Testbed.} To find feasible attack cases in a practical environment, we tried to emulate SD-WAN as similar to real distributed networks as possible for scientific research and run various ONOS applications (e.g., Reactive Forwarding, OpenFlow Driver, Access Control, Stats Provider, etc.). To this end, we constructed a private SD-WAN testbed that spans two campus networks and one enterprise network (see Fig.~\ref{fig:environment}). Those networks are physically distant from each other and connected through a VPN. Each site has one or two hardware OpenFlow switches (i.e., Pica 3297, 3290, and EdgeCore AS4610-54T) controlled by a local cluster, which is composed of a controller and arbiter node. In order to consider as many distributed environments as possible and solve the limitation of the hardware testbed (e.g., the number of nodes), we devised a builder that can synthesize cluster environments automatically. Therefore, we can demonstrate the feasibility of attack case studies including 4 nodes to 16 nodes while considering various configurations in our testbed.

\section{Attack Case Studies}
\label{sec:attack_demo}
In this section, we demonstrate comprehensive cases of cluster attacks discovered by \ourtool{} through the evaluation conducted on our private SD-WAN testbed. We constructed a state machine by learning in a real cluster environment based on ONOS/Atomix and considering a variety of state transitions. From the constructed state machine shown in Fig.~\ref{fig:state_machine}, we extracted 258 seed message sequences, where each sequence consists of 97 input alphabets on average ($min$: 1, $max$: 129, $SD$: 41) with Algorithm~\ref{alg:dfs}. We tried to find abnormal behaviors that can be generated from the state machine and thus generated 1,572,940 random message sequences, where each sequence consists of 120 input alphabets on average ($min$: 1, $max$: 134, $SD$: 19) through the methodology mentioned in \S\ref{sec:state_fuzzing}. When injecting the random messages, we analyzed unexpected operations and vulnerabilities depending on the four major components (\S\ref{subsec:protocol}): (i) Distributed Storage, (ii) Leader Election, (iii) Membership Check, and (iv) Mastership Segmentation. As a result, we found 6 attack scenarios determined as vulnerabilities based on the criteria in \S\ref{subsec:criteria}.

\noindent\textbf{Threat Model.} \ourtool{} is a testing tool that allows network operators to investigate potential attacks against distributed controllers aimed to disrupt or poison the controllers' cluster. We consider an adversary who can only inject \emph{a small number of messages} into the controllers' cluster to conduct attacks without raising suspicion (i.e., in a stealthy manner). This assumption is important because if the network operators detect an ongoing attack against one of the controllers within the cluster, they could quickly disconnect the affected controller from the cluster before the attack is completed. To inject messages into the controllers' cluster, the adversary can either control (i) one of the nodes within the cluster or (ii) any middlebox that receives and analyzes the messages exchanged between controllers. Past work has already demonstrated that SDN controllers are complex components that often contain serious vulnerabilities which can be exploited from both the data plane~\cite{xiao2020unexpected,ujcich2020automated} and the application plane~\cite{ujcich2018cross,lee2016smaller,yoon2017flow}.

\subsection{Distributed Storage}

\begin{figure}[t]
    \centering
    
    \subfloat{
        \includegraphics[width=0.95\linewidth]{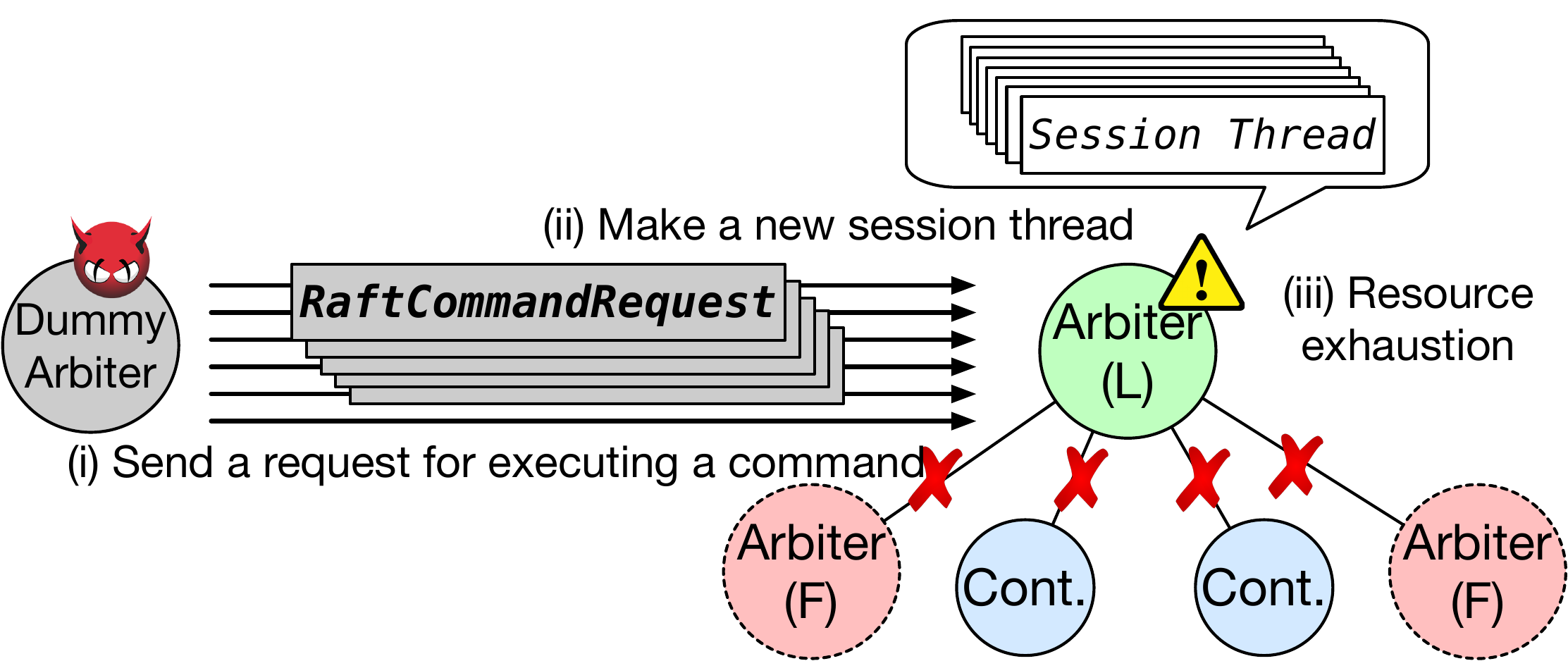}
    }
    
    \subfloat{
        \includegraphics[width=\linewidth]{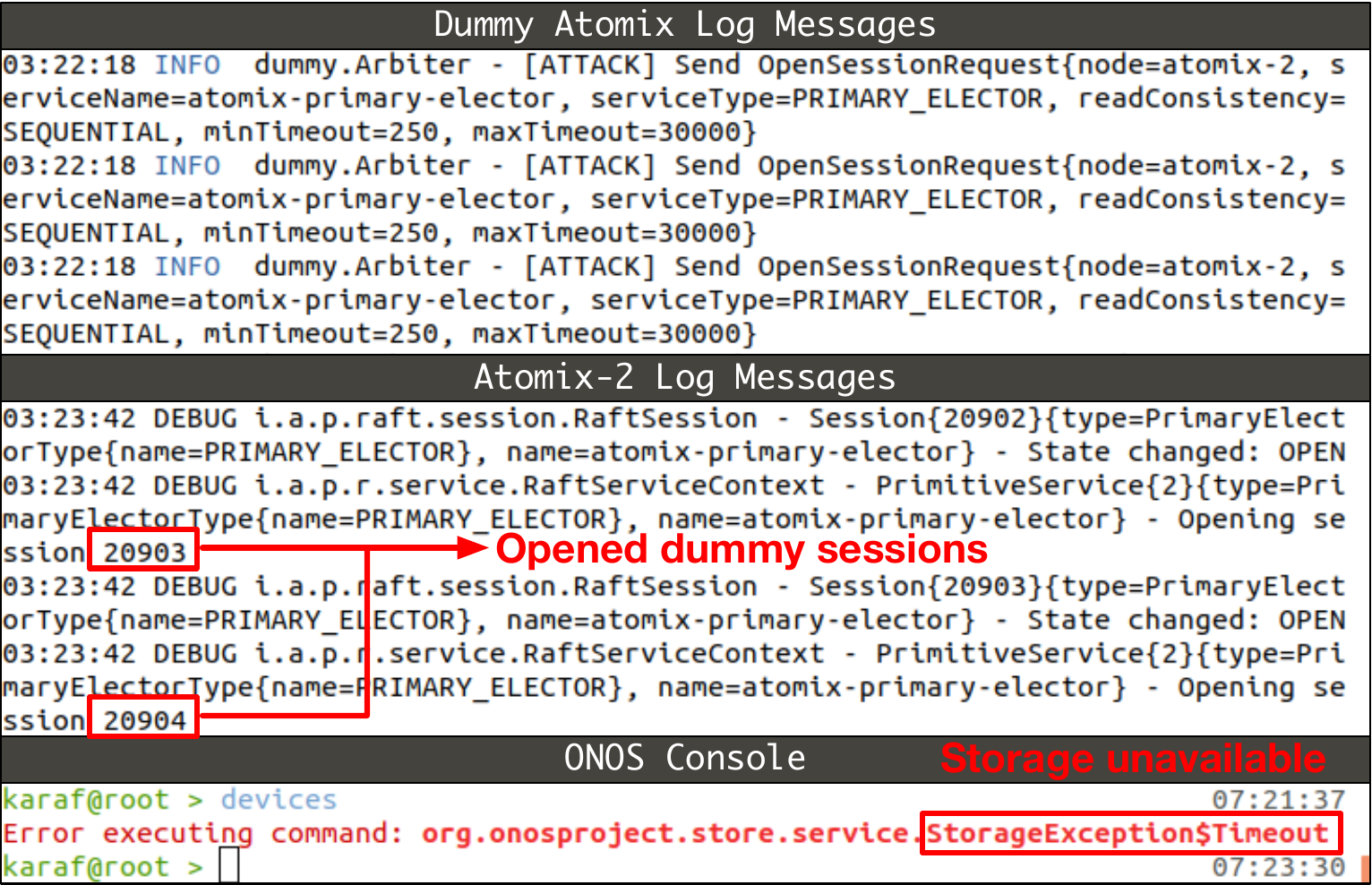}
    }
    \caption{The scenario (top) and result (bottom) of the cluster session flooding attack.}
    \label{fig:session_flooding}
    \vspace{-0.15in}
\end{figure}

\noindent\textbf{Cluster Session Flooding (CVE-2020-35210).} We discover that an adversary can make the target cluster unavailable by performing denial-of-service (DoS) attacks using cluster messages. They can indirectly exhaust the resources of a machine that runs arbiter instances by generating a bunch of cluster session requests (see Fig.~\ref{fig:session_flooding} (top)). (i) They employ dummy nodes that send the leader many \emph{RaftCommandRequest} messages. When creating a message, they use a randomized member ID and network identifier to force the leader to maintain more session states. (ii) When the leader receives a \emph{RaftCommandRequest} message, it creates a dedicated server-side thread to maintain the session state and handle subsequent cluster messages (i.e., the transition V7$\rightarrow$V4 in Fig.~\ref{fig:state_machine}). (iii) The arbiter node is supposed to disconnect the session, which does not receive further messages. However, if the flooding rate exceeds the disconnecting rate, the enormous threads significantly consume the leader's resources. Ultimately, the victim leader node cannot process other essential jobs, such as state replication invoked by the controller nodes. Moreover, it will finally freeze the entire cluster operation.

Fig.~\ref{fig:session_flooding} (bottom) demonstrates this attack scenario against ONOS. A dummy arbiter node continuously sends a bunch of \emph{SessionRequest} messages to the leader arbiter node \texttt{Atomix-2}, which causes the leader node to generate many unnecessary cluster sessions. Subsequently, it turns out that receiving device information from distributed storage is no longer possible, as shown in the console. Therefore, it is determined that this vulnerability is correlated with the criterion ``Excessive Resource Usage'' (DC5 in \S\ref{subsec:criteria}).

\begin{figure}[t]
    \centering
    \subfloat{
        \includegraphics[width=0.95\linewidth]{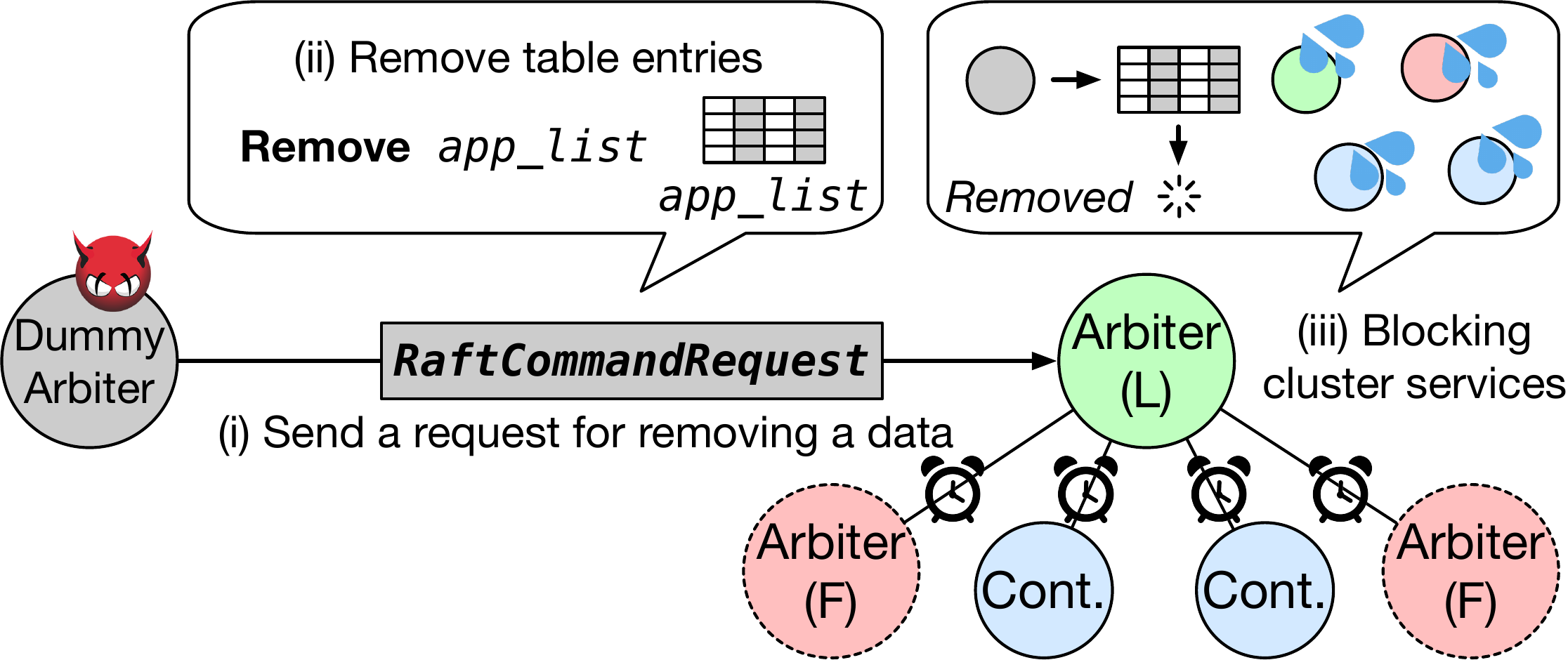}
    }
    
    \subfloat{
        \includegraphics[width=\linewidth]{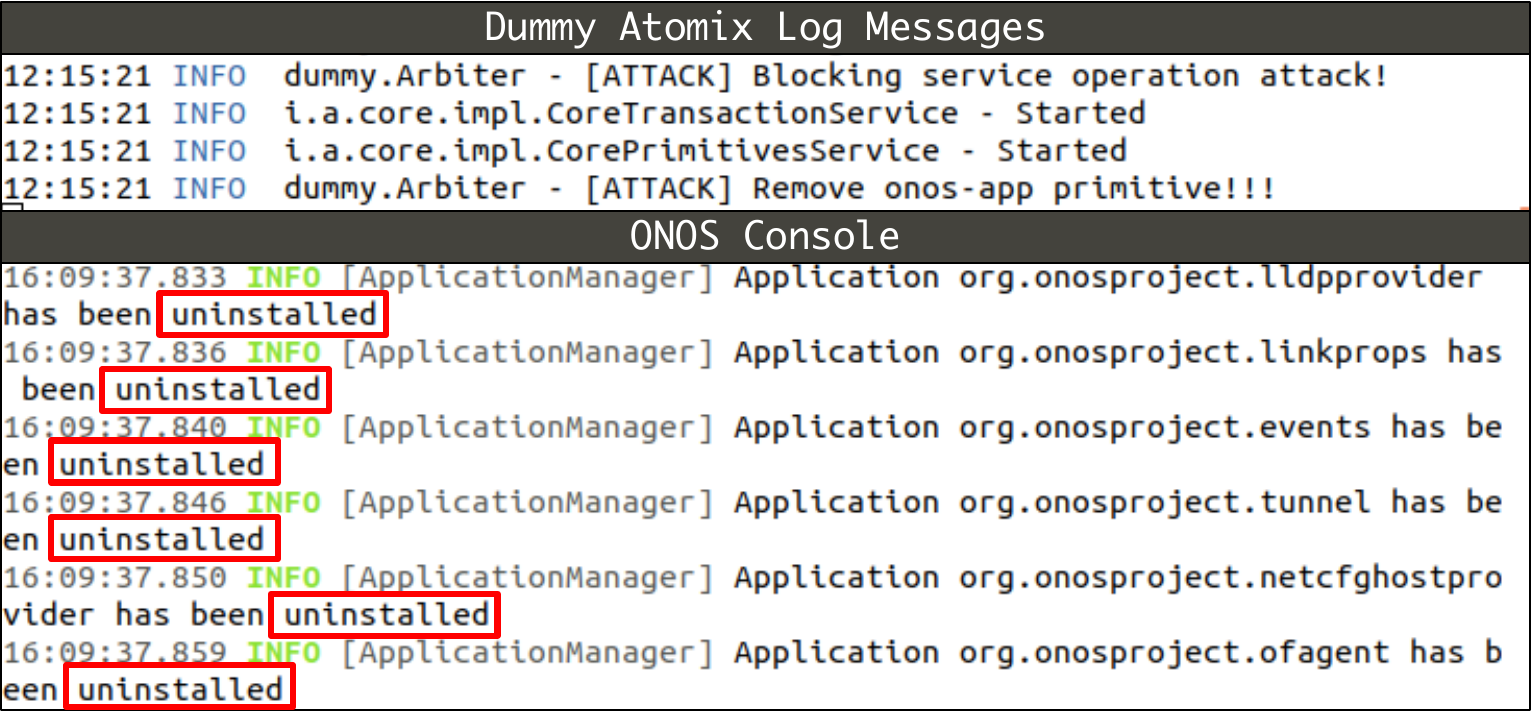}
    }
    \caption{The scenario (top) and result (bottom) of the blocking service operation attack.}
    \label{fig:blocking}
    \vspace{-0.15in}
\end{figure}

\noindent\textbf{Blocking Service Operation (CVE-2020-35214).} Management of synchronized states is essential for maintaining a logically centralized view of controllers. For this, distributed controllers use \emph{shared data} that stores current network states and configuration information with diverse data structures (e.g., map, set, list, counter). Controller applications can write/update/query the data stored in distributed storage, and then the data is synchronized over all nodes in the cluster. Here, an adversary can abuse this propagation mechanism to disrupt entire cluster management: (i) They can invoke an instruction that \texttt{removes} all entries by targeting one of the important primitive tables for network management, such as an application list (i.e., the transition V7$\rightarrow$V4 in Fig.~\ref{fig:state_machine}). (ii) This operation is synchronized over all nodes, removing all installed applications in their local storage. (iii) As a result, benign controller nodes cannot properly use the deleted applications anymore. Fig.~\ref{fig:blocking} (top) elaborates on this attack scenario.

Fig.~\ref{fig:blocking} (bottom) shows the result of the operation abuse attack in ONOS. Once a dummy arbiter node joins a target cluster, it can access the \texttt{onos-apps} map that contains all currently installed SD-WAN applications. When installing and activating an application, controller nodes update the table and synchronize it with each other. The dummy node can call the \texttt{clear} API that removes all application lists of the target primitive. As shown in the console, the controller node regards all applications as being removed, making services unavailable. Consequently, this attack corresponds to the criterion ``Cluster Status Changes'' (DC3 in \S\ref{subsec:criteria}).

\subsection{Leader Election}

\begin{figure}[t]
    \centering
    \subfloat{
        \includegraphics[width=0.95\linewidth]{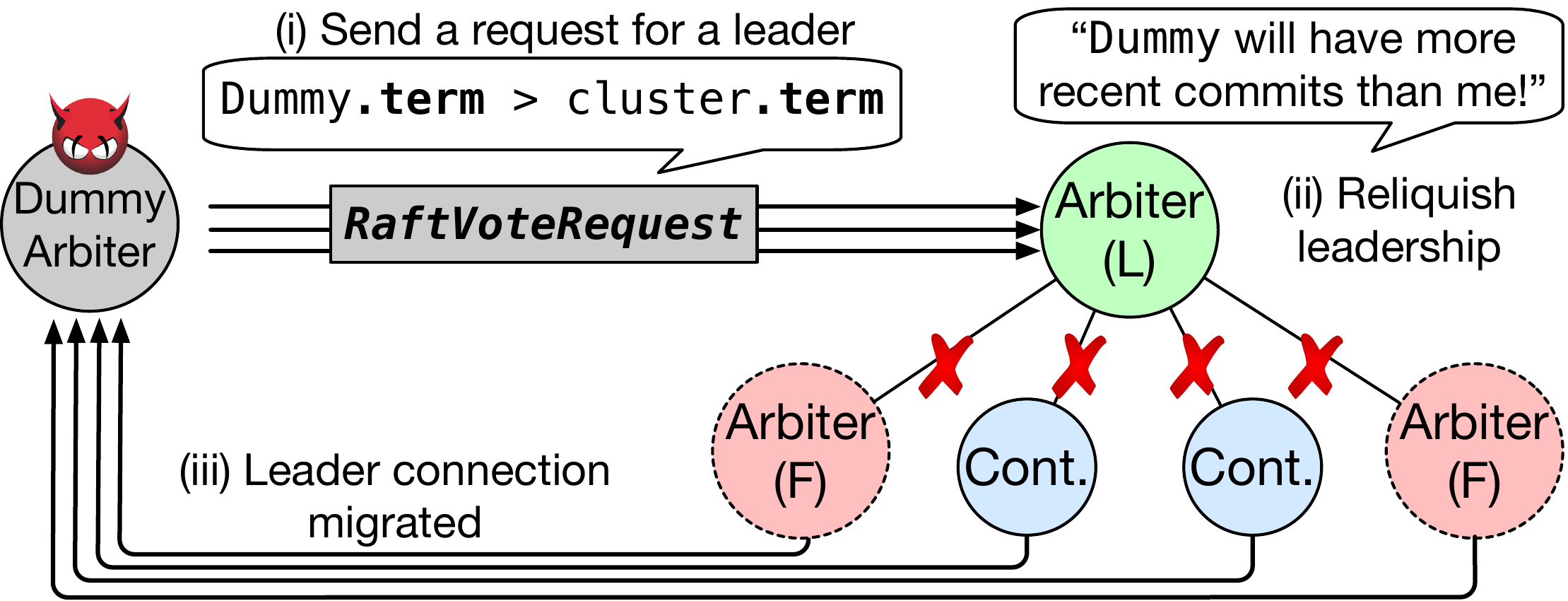}
    }
    
    \subfloat{
        \includegraphics[width=\linewidth]{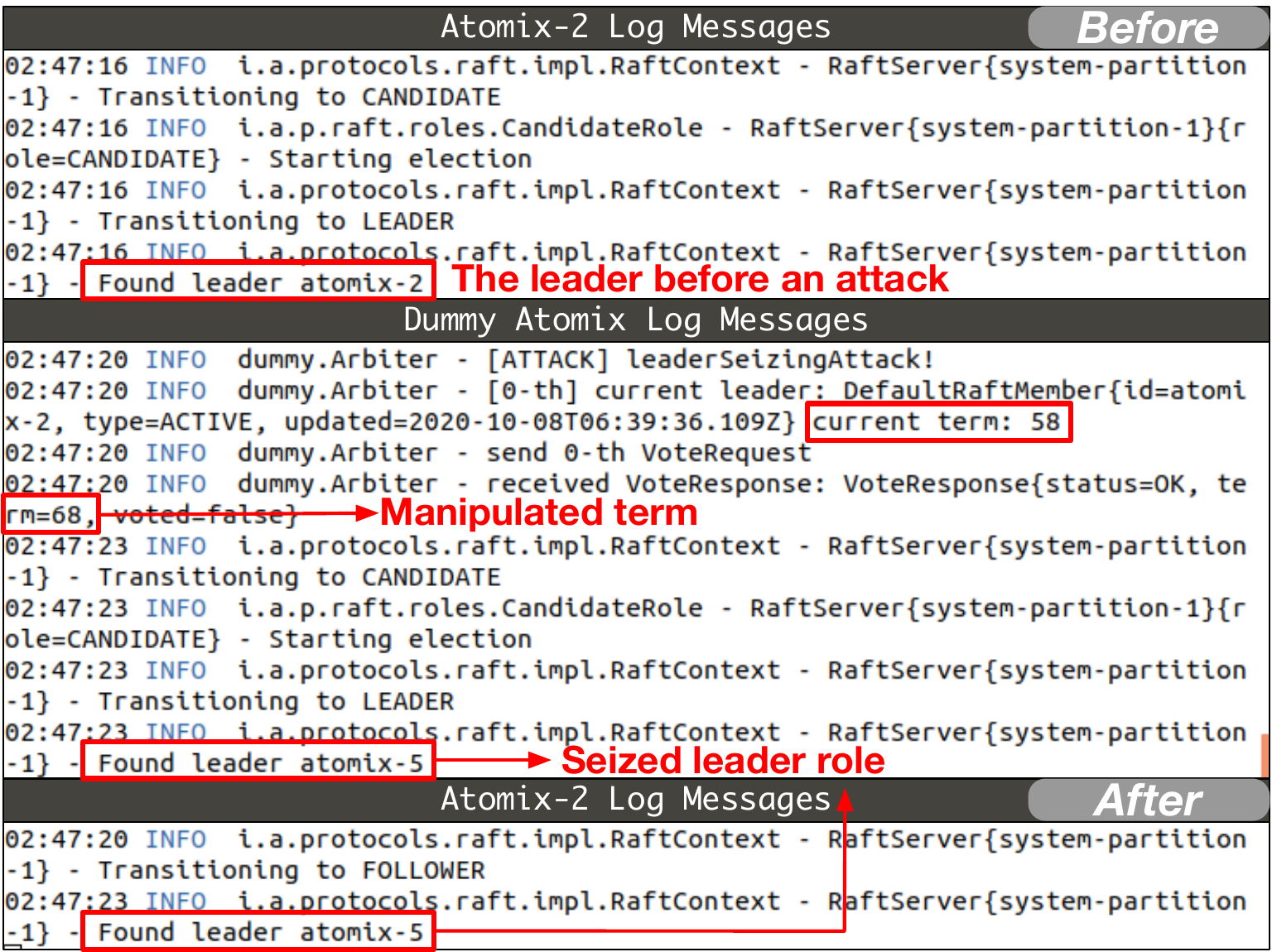}
    }
    \caption{The scenario (top) and result (bottom) of the leader seizing attack.}
    \label{fig:leader_seizing}
    \vspace{-0.15in}
\end{figure}

\noindent\textbf{Seizing Leadership (CVE-2020-35211).} In general, controller nodes maintain \emph{election terms} to track current election period of a cluster (\S\ref{sec:background}). This metadata is to prevent a cluster from suffering a partitioning problem---indicating a situation in which two leaders exist at the same time due to a disconnected link between two partitions---by suppressing a lower-term node to be a leader~\cite{ongaro2014search}. While it is the key design of the Raft algorithm, we discover that an adversary can abuse this to make their dummy node become a new leader by manipulating the election term. Fig.~\ref{fig:leader_seizing} (top) shows how an adversary's node seizes the leadership with an election manipulation process.

Fig.~\ref{fig:leader_seizing} (bottom) illustrates the result of the attack. The dummy node (i.e., \texttt{Atomix-5}) first discovers that the current leader node is \texttt{Atomix-2} (on the top console) before conducting the attack. The dummy node then sends a series of manipulated \emph{RaftVoteRequest} messages that always have a higher election term than the current leader. When receiving those messages, the current leader resigns from the leader role and becomes a follower, and the entire cluster starts a re-election process (i.e., the transition V4$\rightarrow$V3 in Fig.~\ref{fig:state_machine}). Here, the dummy node repeatedly sends the messages with a higher term to compete with other nodes. Then, those messages with a higher term from the dummy node make other nodes in the candidate state relinquish the election and deprive them of chances to become leaders. This process continues until the dummy node receives the majority of votes, and finally, it obtains leadership and will be able to receive all states sent from cluster nodes. Therefore, this vulnerability is correlated with the criterion ``Cluster Status Changes'' (DC2 in \S\ref{subsec:criteria}).

\subsection{Membership Check}
\label{subsec:membership_check}

\begin{figure}[t]
    \centering
    \subfloat{
        \includegraphics[width=0.95\linewidth]{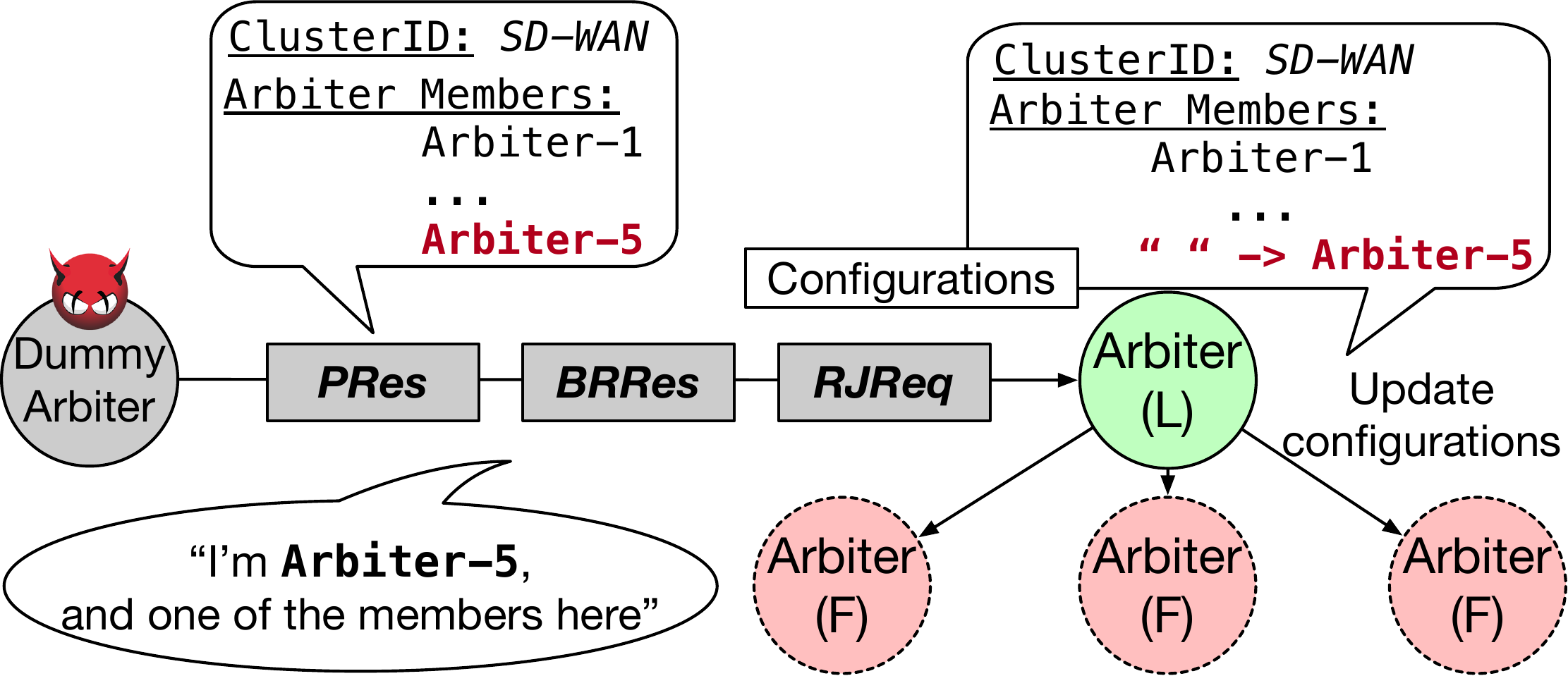}
    }
    
    \subfloat{
        \includegraphics[width=\linewidth]{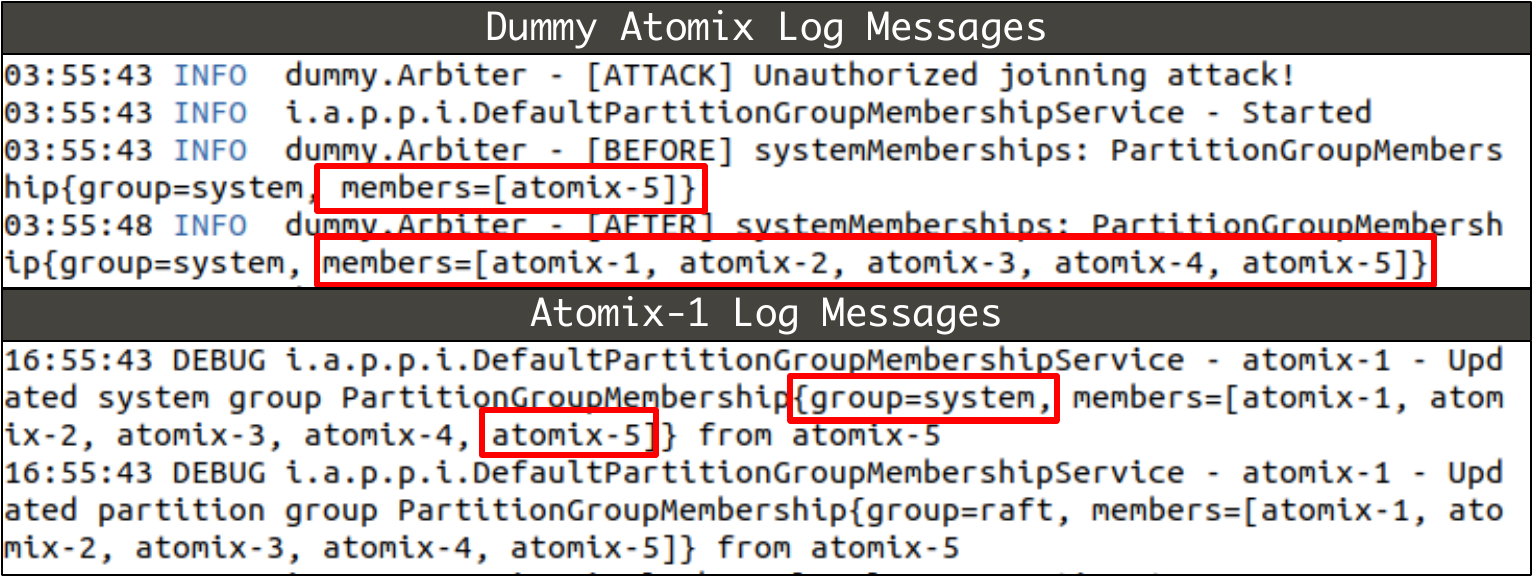}
    }
    \caption{The scenario (top) and result (bottom) of the unauthorized cluster joining attack.}
    \label{fig:join_attack}
    \vspace{-0.15in}
\end{figure}

\noindent\textbf{Unauthorized Cluster Joining (CVE-2020-35209).} Adding/removing a new node to a cluster should be managed carefully because a cluster node can perform crucial operations (e.g., state replication, election). However, we found that an adversary can make their dummy node illegally participate in a target cluster by providing a suitable configuration. Fig.~\ref{fig:join_attack} (top) shows an example attack scenario: When a dummy arbiter node establishes a connection with the cluster, it receives a \emph{BootstrapRequest} message that includes information about a cluster configuration (DC1 in \S\ref{subsec:criteria}). The dummy node then sends a sequence of three messages: \emph{ProbeResponse}, \emph{BootstrapResponse}, and \emph{RaftJoinRequest} to a certain node in the cluster. Here, \emph{BootstrapResponse} contains configuration details specifying the composition of a cluster (e.g., member nodes). Here, the leader node investigates the configuration validity; if not, the message will be rejected. However, a critical security problem is that the cluster allows an unknown node to join the cluster if the configuration includes a correct cluster-ID and a list of existing members (i.e., the transitions V0$\rightarrow$V1$\rightarrow$V4 in Fig.~\ref{fig:state_machine}). For example, if the dummy node sends the cluster-ID (i.e., \texttt{SD-WAN}) with the existing member list (i.e., \texttt{Arbiter-1} to \texttt{4}) including itself, the leader accepts the request and marks the dummy node as a valid one.

Fig.~\ref{fig:join_attack} (bottom) shows the result of the cluster joining attack. Atomix differentiates cluster nodes according to whether a node is in a \emph{system group} or not. As it brings the capability for cluster management, joining the system group empowers an adversary to perform diverse cluster operations. The dummy arbiter node \texttt{Atomix-5} configures itself as one of the system group members to access the core functions of a cluster.

\begin{figure}[t]
    \centering
    
    \subfloat{
        \includegraphics[width=0.95\linewidth]{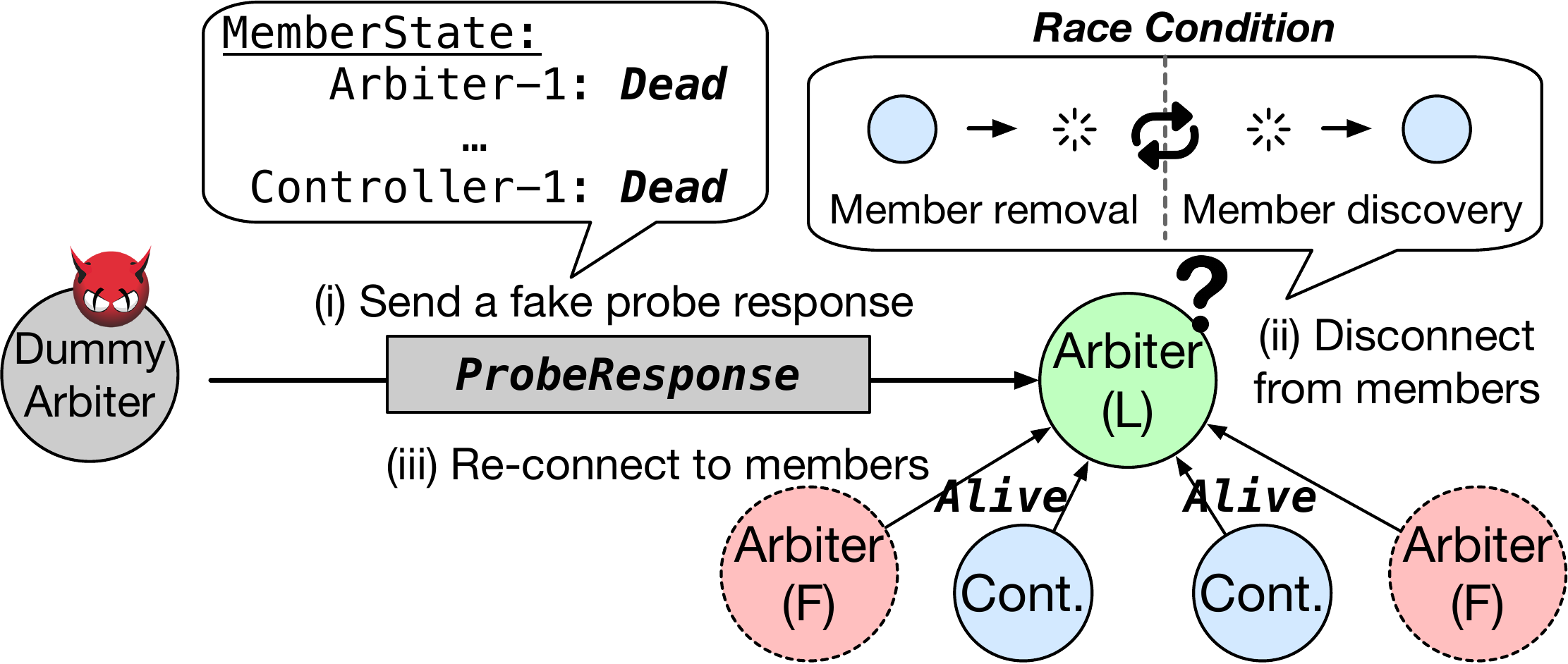}
    }
    
    \subfloat{
        \includegraphics[width=\linewidth]{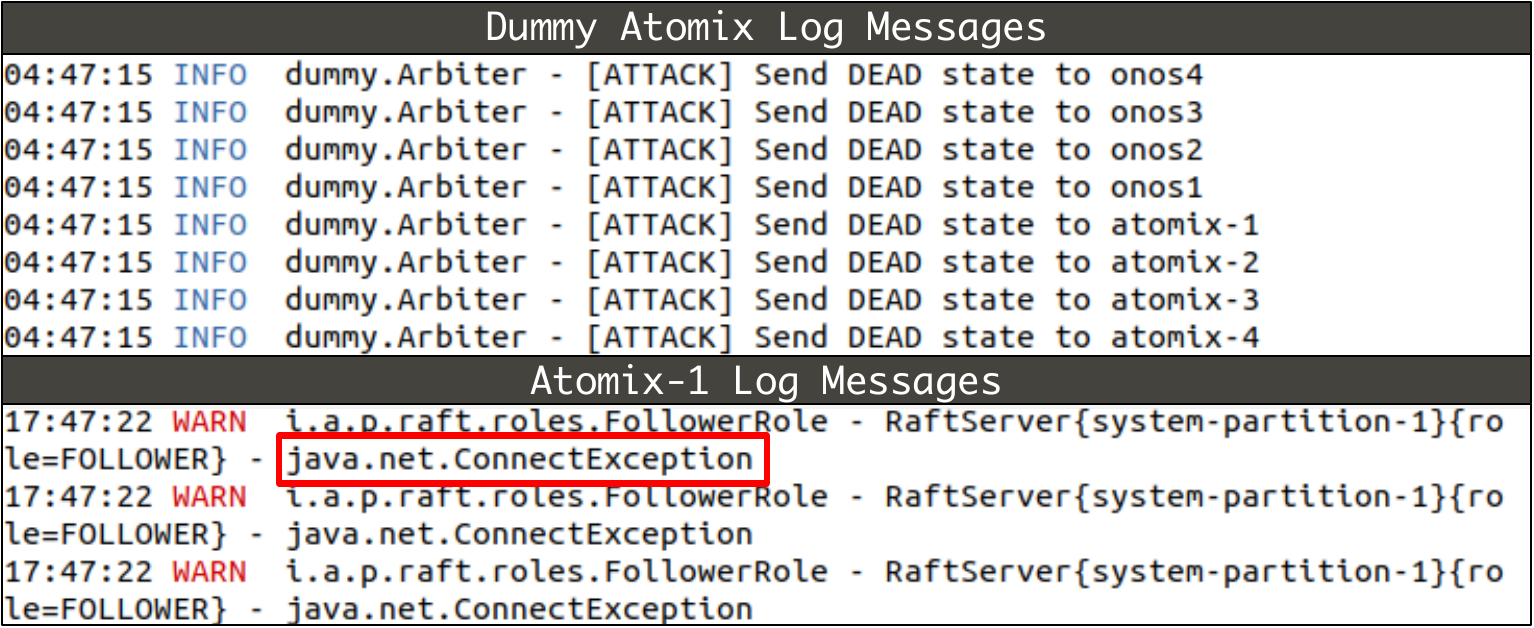}
    }
    \caption{The scenario (top) and result (bottom) of the fake member state advertisement attack.}
    \label{fig:fake_member}
    \vspace{-0.15in}
\end{figure}

\noindent\textbf{Fake Membership State Advertisement (CVE-2020-35216).} SWIM protocol, used in a membership engine, utilizes an \emph{indirect probe} to keep in touch with a node that did not respond to prior direct probes (\S\ref{sec:background}). This feature enables cluster nodes to indirectly obtain the node's aliveness information by synchronizing with its peers (i.e., the transition V2$\rightarrow$V5 in Fig.~\ref{fig:state_machine}). However, we reveal that it is possible to abuse this feature by injecting fake messages that include erroneous membership states. For example, in Fig.~\ref{fig:fake_member} (top), (i) the adversary's dummy node can send fake \emph{ProbeResponse} messages, which tell a lie that all peers are dead, to a victim node. (ii) The victim node then considers this as a truth. Thus, it attempts to execute a series of failure recovery operations, such as removing the dead members from its membership list and re-electing a new leader (if the message contains a current leader). (iii) However, at the same time, since these allegedly ``dead'' nodes are alive, they attempt to synchronize with the victim node to let it know their aliveness. Subsequently, a race condition occurs, which exhausts the victim's resources for updating membership states mainly due to the infinite up and down.

Fig.~\ref{fig:fake_member} (bottom) shows the result evaluated in our environment. When a dummy Atomix node notifies the fake membership state, the \texttt{Atomix-1} immediately removes the members from its membership list. However, spoofed peer members try to connect with the victim node through Raft protocol messages. This race condition makes the victim node's \texttt{RaftServer} refer to the wrong connection information, raising an exception. Therefore, this vulnerability corresponds to the ``Cluster Configuration Changes'' (DC2 in \S\ref{subsec:criteria}).

\begin{figure}[t]
    \centering
    \subfloat{
        \includegraphics[width=0.95\linewidth]{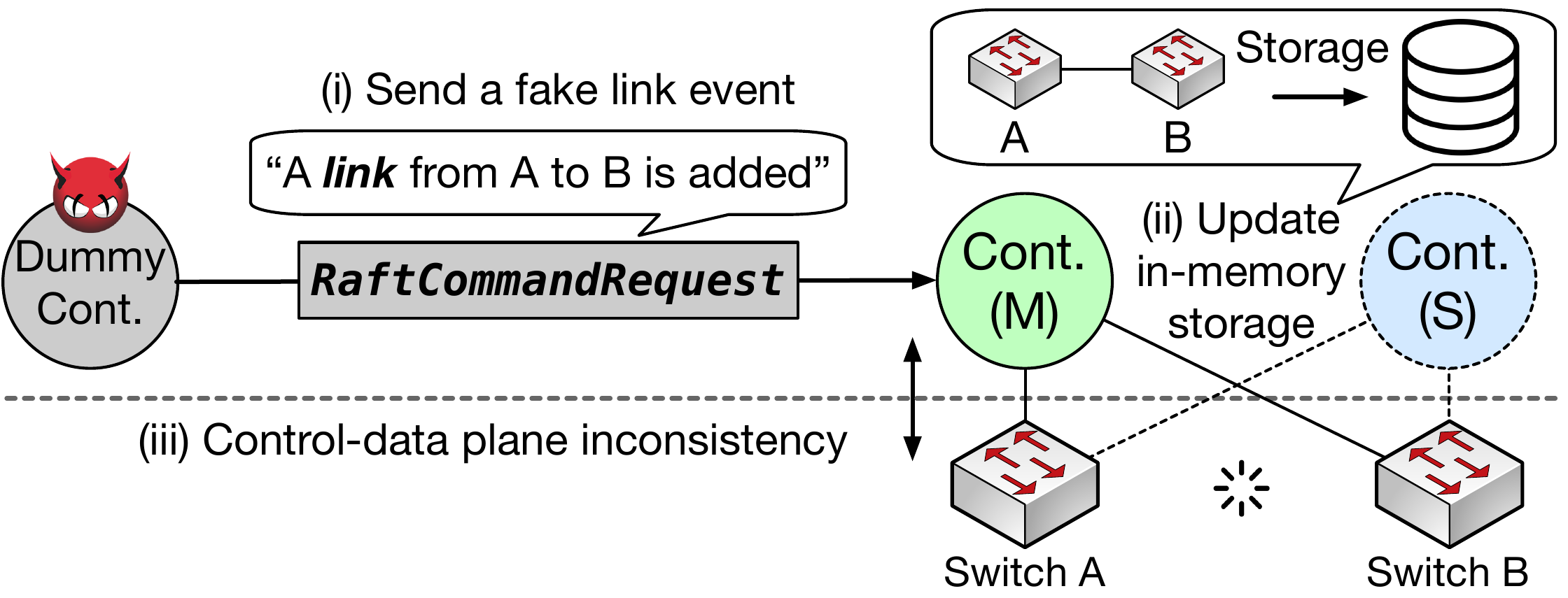}
    }
    
    \subfloat{
        \includegraphics[width=\linewidth]{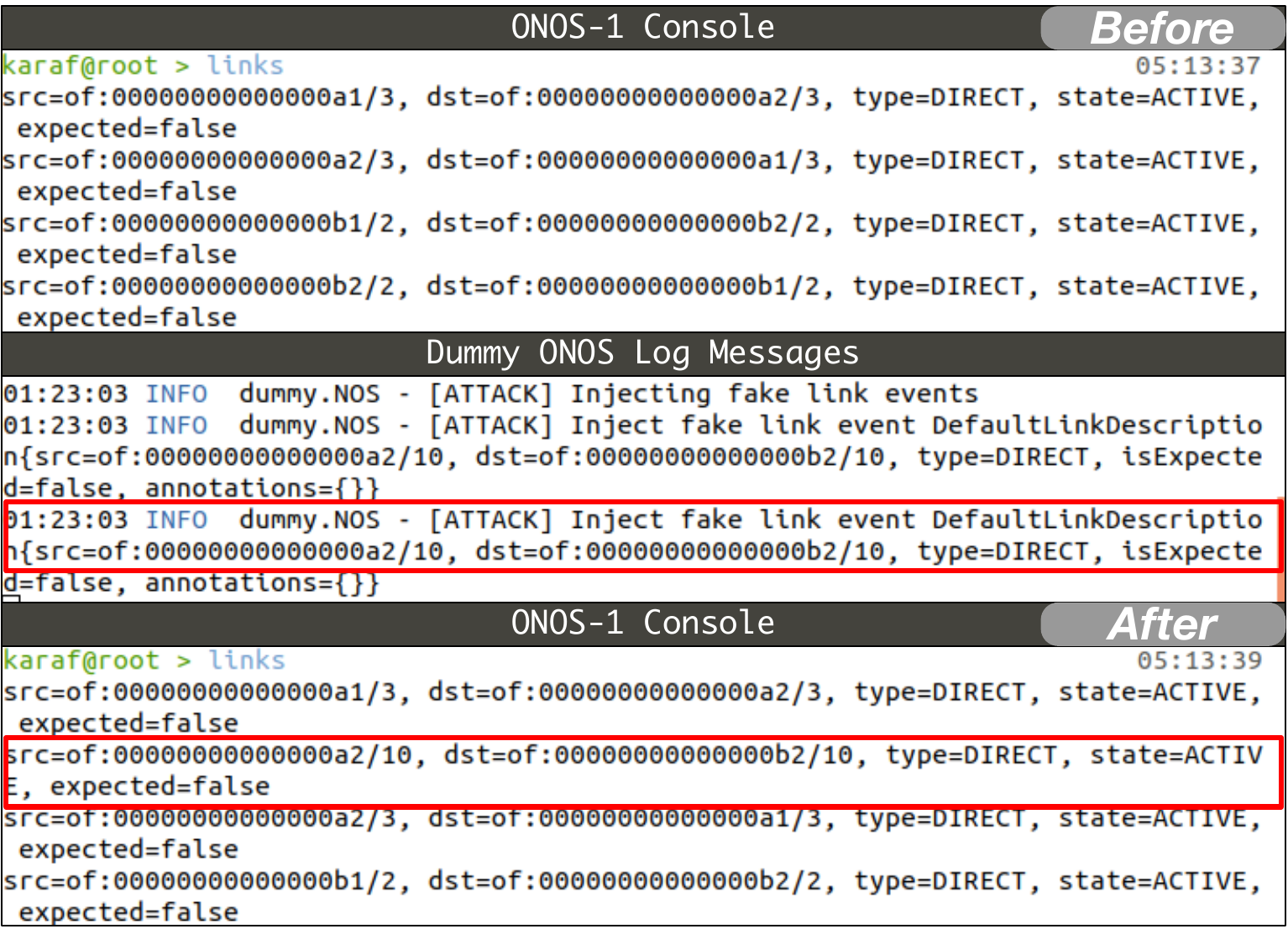}
    }
    \caption{The scenario (top) and result (bottom) of the fake data-plane event injection attack.}
    \label{fig:fake_link}
    \vspace{-0.15in}
\end{figure}

\subsection{Mastership Segmentation}

\noindent\textbf{Fake Data-plane Event Injection (CVE-2020-35213).} A link state is one of the significant assets of an SDN controller to know the current connectivity between switches for determining a correct routing path. For this reason, topology poisoning attacks for SDN controllers have been well-known attack vectors. For example, an adversary injects fake LLDP packets, causing a switch to report a non-existing link to a controller~\cite{hong2015poisoning,skowyra2018effective}. While controllers try to address this with a periodic state-synchronization from switches to a controller, we reveal that such a poisoning attack is also feasible by abusing a cluster relation (see Fig.~\ref{fig:fake_link} (top)). In this scenario, a dummy node crafts and sends a \emph{RaftCommandRequest}, reporting an addition of a fake link that does not exist in the data plane. Then, the master controller node updates its local storage and propagates the event via a distributed storage (i.e., the transition V7$\rightarrow$V4 in Fig~\ref{fig:state_machine}).

Fig.~\ref{fig:fake_link} (bottom) shows the result of this attack in the ONOS controller. The dummy controller node (i.e., \texttt{Dummy ONOS}) sends a fake link event to \texttt{ONOS-1} node, which is the master instance for the switch \texttt{b2}. Before conducting the attack, there is no link between switches \texttt{a2} and \texttt{b2}. However, when we inject a fake event into the master node, it updates its eventually consistent distributed storage without integrity checking. Even worse, we confirm that this poisoned state is not cleaned up until (i) spoofed network devices are re-connected or (ii) the master controller node gets restarted. Note that both cases are time-consuming and cause service disruption in practice. This case corresponds to the criterion ``Network Reachability Changes'' (DC4 in \S\ref{subsec:criteria}).

\subsection{Execution Time Measurement}
\begin{figure}[t]
    \centering
    \includegraphics[width=\linewidth]{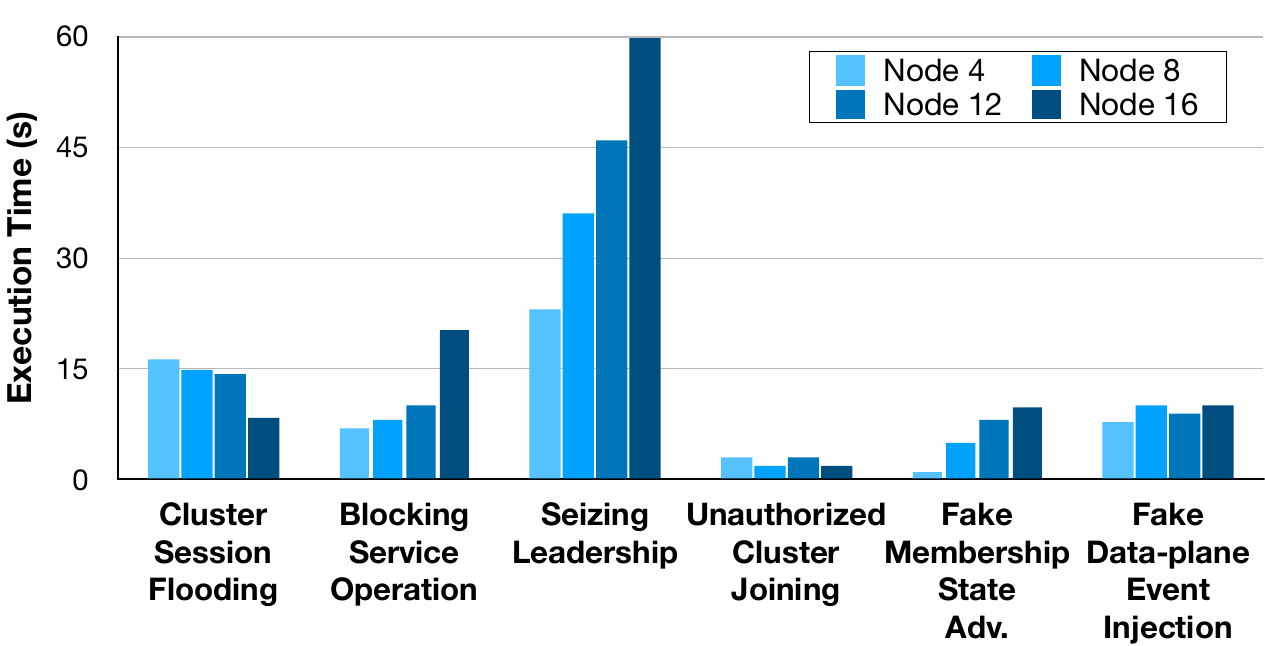}
    \caption{Attack execution time across different cluster sizes.}
    \label{fig:execution_time}
    \vspace{-0.15in}
\end{figure}

Finally, we evaluated the duration of each attack across different cluster sizes to confirm its practical feasibility. Fig.~\ref{fig:execution_time} illustrates the execution time of each attack case in different clusters, including 4 nodes to 16 nodes. In the case of Cluster Session Flooding, the more cluster nodes increase, the better session flooding can be exploited. This result stems from the fact that a larger cluster should manage many service operations and nodes, and thus it can be highly vulnerable to resource exhaustion attacks. Additionally, the Blocking Service Operation and Fake Membership State Advertisement attack show that longer execution time is needed as more nodes are added because the cluster states need to be spread over all controller nodes. Also, the Seizing Leadership attack shows that the execution time gradually increases depending on the cluster size because the execution time can be influenced by the intensity of competition, which is determined by the number of cluster nodes. Compared with the above cases, we figure out that the Unauthorized Cluster Joining attack case maintains constant execution time regardless of the number of cluster nodes because an intruding node already has predefined configuration information and needs to send a join message to one cluster node, i.e., leader. Likewise, the Fake Data-plane Event Injection attack maintains constant execution time because the master node propagates events with broadcasting; thus, the cluster size does not affect execution time.

\section{Discussion and Limitations}
\label{sec:discussion}

This section provides in-depth discussions for analyzing the root cause of vulnerabilities in distributed SDN controllers and proposes countermeasures to prevent them from being abused. We also discuss the limitations associated with \ourtool{}.

\newcolumntype{M}[1]{>{\arraybackslash}m{#1}}
\newcolumntype{C}[1]{>{\centering\arraybackslash}m{#1}}

\begin{table*}[t]
    \centering
    \small
    \caption{Comparison of existing SDN attack detection tools and \ourtool{} (* denotes that the feature is partially supported but not a main goal.).}
    \resizebox{\textwidth}{!}{
    \begin{tabular}{M{3cm} M{7.5cm} M{3cm} M{2.5cm} M{2cm}}
    \toprule
        \textbf{Work} & \textbf{Main Focus} & \textbf{Methodology} & \textbf{Distributed-aware} & \textbf{State-aware} \\ \midrule

         NICE~\cite{canini2012nice} & Faulty SDN applications & Model checking & \xmark & \cmark \\ \midrule

         STS~\cite{scott2014troubleshooting} & Minimal casual sequences for troubleshooting & Delta debugging  & \cmark* & \xmark \\ \midrule

         Jury~\cite{mahajan2016jury} & Faulty controllers in a cluster & Event hooking & \cmark & \xmark \\ \midrule

         ConGuard~\cite{xu2017attacking} & Race conditions on SDN applications & Dynamic analysis  & \xmark & \xmark \\ \midrule
        
         DELTA~\cite{lee2017delta,lee2020comprehensive} & Vulnerabilities in Northbound interfaces and apps & Black-box fuzzing & \xmark & \cmark \\ \midrule
         
         BEADS~\cite{jero2017beads} & Vulnerabilities in Northbound interfaces & Black-box fuzzing & \xmark & \xmark \\ \midrule
         
         ATTAIN~\cite{ujcich2017attain} & Vulnerabilities in controllers & Attack injection & \xmark & \cmark \\ \midrule
         
         AIM-SDN~\cite{dixit2018aim} & Semantic gap in controller datastores & Black-box fuzzing & \xmark & \xmark \\ \midrule
         
         AudiSDN~\cite{lee2020audisdn,lee2022framework} & Policy inconsistencies between a controller and switches & Black-box fuzzing & \xmark & \cmark \\ \midrule

         Spider~\cite{li2022spider} & Stateful performance issues in controllers & Grey-box fuzzing & \xmark & \xmark \\ \midrule

         Intender~\cite{kim2023intender} & Vulnerabilities in intent-based networking (IBN) & Black-box fuzzing & \xmark & \cmark  \\ \midrule
         
         \textbf{\ourtool{} (our work)} & Vulnerabilities in East/Westbound interfaces & Protocol state fuzzing & \cmark & \cmark \\
    \bottomrule
    \end{tabular}
    }
    \label{tab:comparison}
    \vspace{-0.1in}
\end{table*}

\noindent\textbf{Weak authentication for node validity.} While current distributed controllers require a \emph{cluster ID} when a node attempts to join a cluster, we have discovered that adversaries can easily estimate this ID. For example, we found that ONOS uses default string values as cluster IDs, such as \texttt{onos} and \texttt{raft}~\cite{onos_authentication}. This weakness allows adversaries to predict the cluster ID using dictionary attacks based on those strings. To avoid this situation, one may leverage an idea from TCP handshaking. Specifically, distributed controllers can use a random ID to negotiate with cluster members similarly to the TCP random sequence number. This approach makes it difficult for adversaries to guess the required IDs.

\noindent\textbf{Lack of inspection for storage integrity.} Eventual consistency aims to alleviate the constraints of strong consistency, which are not scalable in large-scale networks, and therefore, it is widely adopted in distributed SDN controllers~\cite{berde2014onos}. However, we demonstrate that a storage system built on eventual consistency is not easily cleaned if it becomes tainted with fake information (i.e., Fake Data-plane Event Injection). The underlying issue is that the storage system updates its entries in a notification-based manner. For instance, in ONOS, the storage system does not eliminate the contaminated entry if there is no \emph{LinkRemoved} event to signal to remove a fake link. As a solution, we propose that distributed controllers should periodically cross-reference the events injected from peer nodes with actual data-plane information.

\noindent\textbf{Architectural issues in the decoupled structure.} Although the recent architecture of distributed controllers---where certain functionalities are separated as arbiters (\S\ref{sec:exp})---offers benefits in terms of flexibility, we argue that it significantly expands attack surfaces for adversaries. For instance, due to the separation of distributed modules from ONOS, Atomix exposes a vulnerability that allows unknown nodes to join a cluster even if they are not registered as initial members (\S\ref{subsec:membership_check}). To address this, the controller vendor should provide an alternative that integrates the controllers with arbiters to reduce attack surfaces.

\noindent\textbf{Lack of access control for distributed controllers.} We observe that current distributed controllers do not mandate authentication when nodes invoke APIs to perform cluster operations, granting adversaries access to various harmful attack scenarios. Specifically, we propose that crucial cluster operations (e.g., adding new nodes, removing applications) should require additional authentication when executed. Furthermore, API-level access control should be implemented to prevent malicious nodes from exploitation. However, despite the existence of access control models for single controllers~\cite{yoon2017security,porras2015securing,fortnox,shin2014rosemary}, none have addressed a permission model for distributed controllers.

\noindent\textbf{Obtaining input/output alphabets via manual analysis.} One limitation of \ourtool{} is that it relies on the manual analysis of source code to obtain input/output alphabets of distributed SDN controllers. The primary reason is that there is no standard specification for East-West interfaces, in contrast to TLS/DTLS cases~\cite{de2015protocol,fiterau2020analysis}. However, we argue that those efforts are minimal as the alphabets can be easily obtained from their public code base (e.g., ONOS~\cite{atomix_cluster_code}, ODL~\cite{odl_cluster_code}).

\noindent\textbf{Supporting other controllers in \ourtool{}.} The current \ourtool{} prototype uses the ONOS controller as a representative example. However, it is important to note that \ourtool{} can be extended to test other controllers with minimal effort, such as ODL. For example, ODL also relies on the Raft algorithm for consensus; thus, many input alphabets of ONOS can be utilized. While this requires us to implement an additional proxy (see Fig.~\ref{fig:system_overview}) that can generate ODL-specific messages, it could be achieved by leveraging the existing code base. For example, ODL is an open-source distributed controller whose cluster message implementations are publicly available~\cite{odl_cluster_code}.


\noindent\textbf{Missing edge cases by the simplified model.} Since \ourtool{} takes a node-to-cluster model to reduce large state space, some edge cases can be missing in the cluster communication. We perform a manual analysis for the extracted state machine by comparing it with the controller's source code to avoid excluding important states or transitions. We confirmed that most of them are incorporated, and excluded cases do not play an essential role in cluster operations. While this manual analysis violates our design consideration that seeks an automatic tool, it is adopted in other protocol state fuzzing tools~\cite{de2015protocol,fiterau2020analysis}.

\section{Related Work}
\label{sec:related_work}

\noindent\textbf{Vulnerabilities in SDN.} Security concerns have been raised over various SDN components since its inception~\cite{yoon2017flow,scott2013sdn,kreutz2013towards,kim2024enhancing}. For example, the SDN centralized architecture is fundamentally weak to control-plane saturation when a switch generates a huge number of flow requests, causing \emph{PacketIn} flooding attacks~\cite{wang2015floodguard,shin2013attacking,curtis2011devoflow,shin2013avant}. While this is a well-known attack vector in a single controller, no previous works studied flooding attacks in distributed controllers, as demonstrated in our paper. On the other hand, a malicious application can execute harmful operations due to the lack of permissions in Northbound interfaces~\cite{lee2016smaller,shin2014rosemary,ropke2015sdn}. However, prior studies did not focus on the security problem of East-West interfaces. In addition, controller storage can be corrupted by malicious data-plane events~\cite{hong2015poisoning,skowyra2018effective,dhawan2015sphinx,jero2017identifier,dixit2018aim,xu2017attacking,marin2019depth} or by abusing a Northbound interface~\cite{lee2020audisdn,lee2022framework}, causing inconsistencies between control and data plane. Distinguished by the previous attacks, we discover a poisoning attack via East-West interfaces. As such, our work is the first to investigate the vulnerabilities of East-West interfaces in distributed controllers comprehensively.

\noindent\textbf{Securing SDN controllers.} There have been efforts towards building secure SDN controllers resilient to known SDN vulnerabilities. Role-based access controls (RBAC) have been regarded as promising solutions for preventing malicious API invocation~\cite{yoon2017security,porras2015securing,wen2016sdnshield}. Meanwhile, secure controller architectures have been proposed with modular approaches~\cite{shin2014rosemary,chandrasekaran2016isolating,narm2018barista}. While these works contributed to building secure architectures in a single controller case, the design of secure architecture for distributed controllers should be studied according to the disclosed vulnerabilities in our paper.

\noindent\textbf{SDN attack detection tools.} To find potential vulnerabilities in SDN, many attack detection tools have been proposed with various techniques. NICE~\cite{canini2012nice} utilized model checking and symbolic execution to find bugs in SDN applications while reducing the large state space. STS~\cite{scott2013sdn} adopted delta debugging to identify minimal causal sequences for troubleshooting issues within SDN environments. ConGuard~\cite{xu2017attacking} used dynamic analysis based on happens-before relationships to discover harmful race conditions in SDN applications. DELTA~\cite{lee2017delta,lee2020comprehensive} and BEADS~\cite{jero2017beads} used black-box fuzzing to find vulnerabilities in SDN northbound interfaces and applications. ATTAIN~\cite{ujcich2017attain} proposed an attack injection framework to discover vulnerabilities in controllers. AIM-SDN~\cite{dixit2018aim} found semantic gaps in controller datastores using black-box fuzzing. AudiSDN~\cite{lee2020audisdn} discovered policy inconsistencies with black-box fuzzing. Spider~\cite{li2022spider} used grey-box fuzzing to locate stateful performance issues in controllers. Intender~\cite{kim2023intender} aimed to find vulnerabilities within the intent-based networking subsystem in SDN controllers through black-box fuzzing.

However, none of the existing works focused on the vulnerabilities of the East-West interfaces in distributed controllers, which is the contribution of \ourtool{}. The most similar one to ours is Jury~\cite{mahajan2016jury}, which aimed to pinpoint faulty controllers in an SDN cluster. However, it did not consider attack scenarios that abuse the protocols used in distributed controllers. Also, while we observe that several fuzzing tools utilized a state machine, most relied on manual analysis of specification or source code, which is difficult to achieve in distributed controllers. In contrast, \ourtool{} learns a state machine using automata learning and generates a single yet relatively simple one via pruning methodologies. Table~\ref{tab:comparison} shows the comparison of existing tools and \ourtool{}.

\section{Conclusion}
\label{sec:conclusion}

Distributed SDN controllers have received significant attention from industry and academia to realize more flexible and efficient wide-area networking environments (i.e., SD-WAN). However, increasing SD-WAN usage attracts adversaries since successful attacks against distributed controllers guarantee control over more critical networks. Thus, verifying the security issues in distributed controllers is indispensable. For this reason, we propose an automatic testing tool, \ourtool{}, for systematically learning states in an SDN cluster and conducting state-aware fuzzing. We show that \ourtool{} allows us to discover unknown vulnerabilities from a popular SDN controller, ONOS. To our knowledge, this is the first work to automatically find vulnerabilities in a distributed controller in an SD-WAN environment. By utilizing \ourtool{}, network operators can effectively and efficiently conduct in-depth testing to uncover any unknown vulnerabilities in distributed controllers. We believe that our work assists researchers in discovering more possible vulnerabilities in SD-WAN.

\section*{Acknowledgment}

We extend our gratitude to Youngjin Jin for his initial proofreading efforts, to Kwanwoo Kim for naming our system \ourtool{}, and to Prof. Kyungbaek Kim for generously providing access to his network, which facilitated the setup of our SD-WAN testbed. The present research has been conducted by the Research Grant of Kwangwoon University in 2022. Also, this work was partially supported by the National Research Foundation of Korea (NRF) grant funded by the Korean government (MSIT) (No. RS-2022-00166401) and by the European Union’s Horizon Europe project under grant agreements No. 101070473 (FLUIDOS) and No. 101092950 (EDGELESS).

\bibliography{references}
\bibliographystyle{IEEEtran}
\vspace{-0.20in}


\newpage

\begin{IEEEbiography}[{\includegraphics[width=1in,height=1.25in,clip,keepaspectratio]{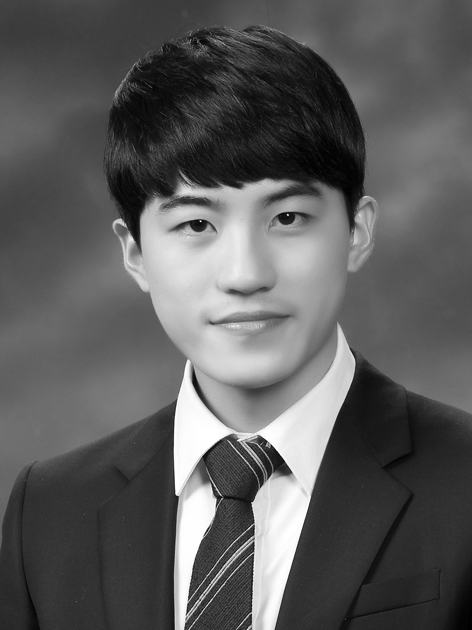}}]
{Jinwoo Kim} is an Assistant Professor in the School of Software at Kwangwoon University, Seoul, South Korea. He received his Ph.D. from the School of Electrical Engineering, his M.S. degree from the Graduate School of Information Security from KAIST, and his B.S. degree from Chungnam National University in Computer Science and Engineering. His research topic focuses on investigating security issues with software-defined networks and cloud systems.
\end{IEEEbiography}
\vspace{-0.5in}

\begin{IEEEbiography}[{\includegraphics[width=1in,height=1.25in,clip,keepaspectratio]{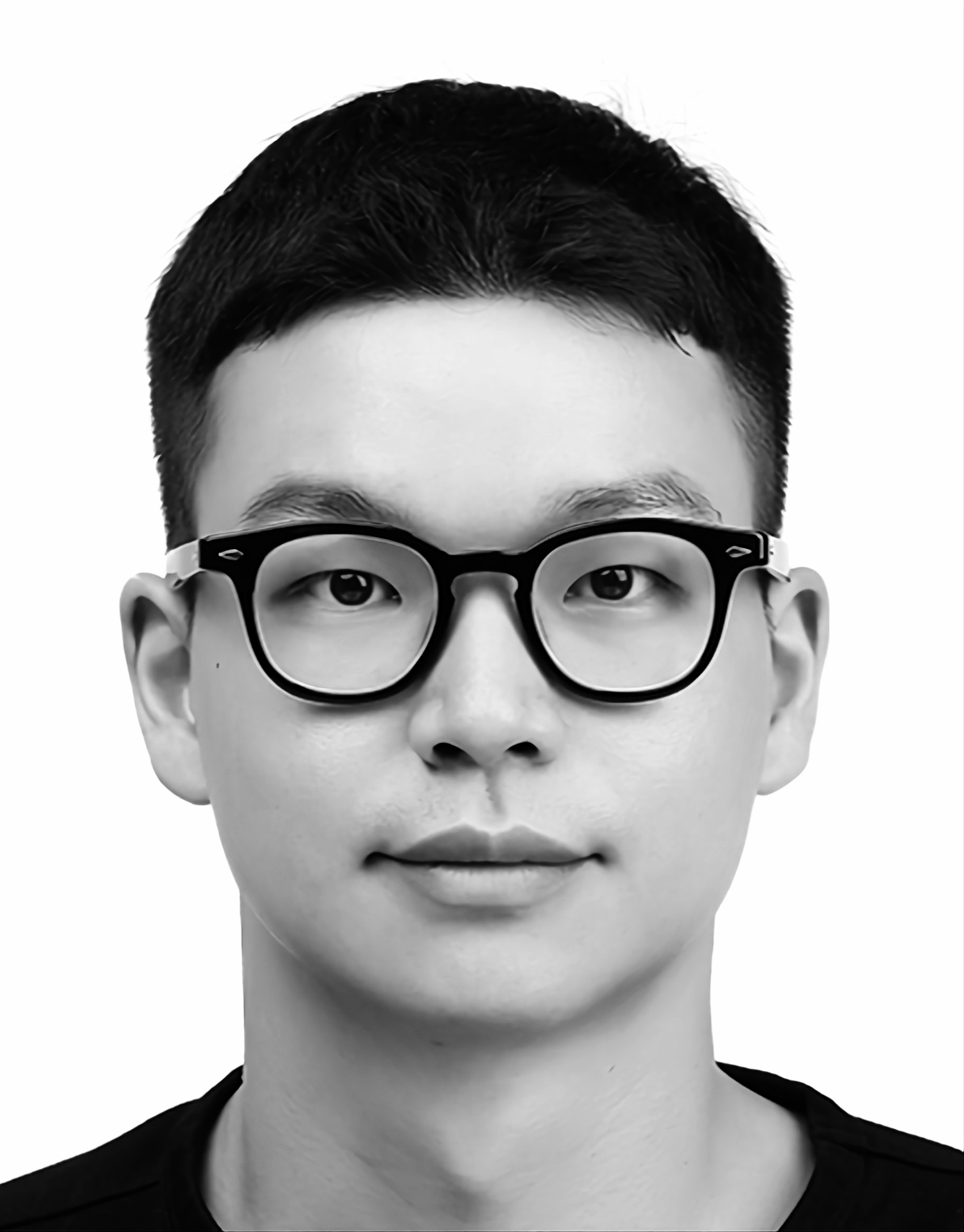}}]
{Minjae Seo} is a Researcher at ETRI, Daejeon, South Korea. He received his M.S. degree from the Graduate School of Information Security at KAIST and his B.S. degree in Computer Engineering from Mississippi State University. His research interests include Software-defined networking security, network fingerprinting, and deep learning-based network systems.
\end{IEEEbiography}
\vspace{-0.5in}

\begin{IEEEbiography}[{\includegraphics[width=1in,height=1.25in,clip,keepaspectratio]{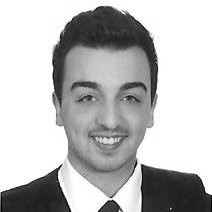}}]
{Eduard Marin} is a Research Scientist at Telefonica Research, Spain. He received his PhD from KU Leuven, Belgium, and his B.S. and M.S. in Telecommunications Engineering from UPC, Spain. After obtaining his PhD, he was a visiting researcher at the University of Padua (Italy) and a postdoctoral researcher at the University of Birmingham (UK). His main research interests lie at the intersection of Networks, Systems, and Security. In particular, he is interested in topics such as Software Defined Networking (SDN), programmable data planes, cloud computing security and cyber-threat intelligence.
\end{IEEEbiography}
\vspace{-0.5in}

\begin{IEEEbiography}[{\includegraphics[width=1in,height=1.25in,clip,keepaspectratio]{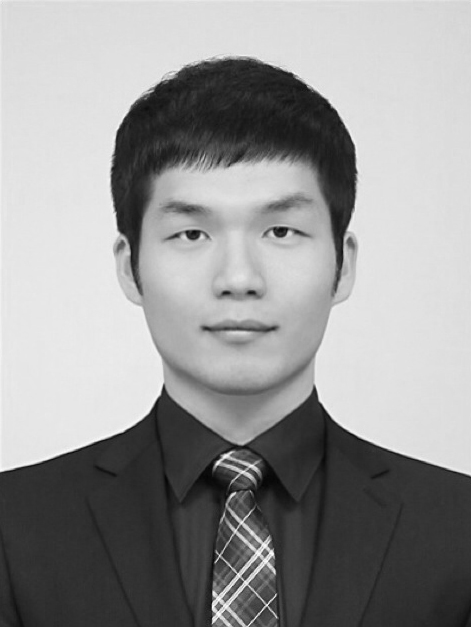}}]
{Seungsoo Lee} is an Assistant Professor in the Department of Computer Science and Engineering at Incheon National University, Incheon, South Korea. He received his B.S. degree in Computer Science from Soongsil University in Korea. He received his Ph.D. and M.S. degrees in Information Security from KAIST. His research interests focus on cloud computing and network systems security. He mainly focuses on software-defined networking (SDN), network function virtualization (NFV), containers, and security issues.
\end{IEEEbiography}
\vspace{-0.5in}

\begin{IEEEbiography}[{\includegraphics[width=1in,height=1.25in,clip,keepaspectratio]{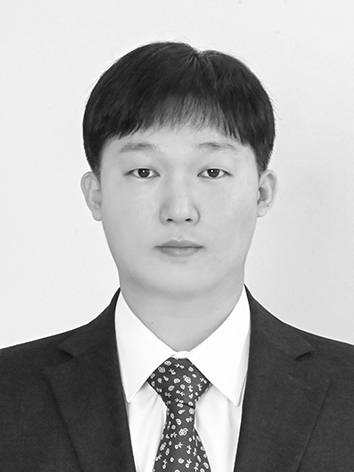}}]
{Jaehyun Nam} is an Assistant Professor at the Department of Computer Engineering, Dankook University, South Korea. He received his Ph.D. and M.S. in School of Computing (Information Security) from KAIST and his B.S. in Computer Science and Engineering from Sogang University in Korea. His research interests focus on networked systems and security. He is especially interested in performance and security issues in cloud and edge computing systems, including SDN/NFV, IoT, and containers.
\end{IEEEbiography}
\vspace{-0.5in}

\begin{IEEEbiography}[{\includegraphics[width=1in,height=1.25in,clip,keepaspectratio]{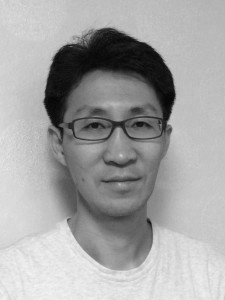}}]
{Seungwon Shin} is an Associate Professor in the School of Electrical Engineering at KAIST and an Executive Vice President at Samsung Electronics. He received his Ph.D. in Computer Engineering from the Department of Electrical and Computer Engineering at Texas A\&M University and his M.S. and B.S. degrees from KAIST, both in Electrical and Computer Engineering. His research interests span the areas of Software-defined networking security, IoT security, Botnet analysis/detection, dark web analysis, and cyber threat intelligence.
\end{IEEEbiography}

\end{document}